# The Energy-Momentum Tensor in General Relativity and in Alternative Theories of Gravitation, and the Gravitational vs. Inertial Mass

### Hans C. Ohanian

*Department of Physics*
*University of Vermont, Burlington, VT 05405-0125*

**Abstract.** We establish a general relation between the canonical energy-momentum tensor of Lagrangian dynamics and the tensor that acts as the source of the gravitational field in Einstein's equations, and we show that there is a discrepancy between these tensors when there are direct nonminimal couplings between matter and the Riemann tensor. Despite this discrepancy, we give a general proof of the exact equality of the gravitational and inertial masses for any arbitrary system of matter and gravitational fields, even in the presence of nonminimal second-derivative couplings and/or linear or nonlinear second-derivative terms of any kind in the Lagrangian. The gravitational mass is defined by the asymptotic Newtonian potential at large distance from the system, and the inertial mass is defined by the volume integral of the energy density determined from the canonical energy-momentum tensor. In the Brans-Dicke scalar field theory, we establish that the nonminimal coupling and long range of the scalar field leads to an inequality between the gravitational and inertial masses, and we derive an exact formula for this inequality and confirm that it is approximately proportional to the gravitational self-energy (the Nordvedt effect), but with a constant of proportionality different from what is claimed in the published literature in calculations based on the PPN scheme. Similar inequalities of gravitational and inertial masses are expected to occur in other scalar and vector theories.

PACS numbers: 04.20.-q, 04.20.Cv, 04.20.Fy, 04.25.Nx, 11.15.Kc

## I. INTRODUCTION

In Newton's theory of gravitation, the ratio of gravitational mass to inertial mass is assumed to be a universal constant, the same for all bodies, and we can adopt a value of 1 for this ratio, so that the inertial mass becomes exactly equal to the gravitational mass, $M_G = M_I$. The universality of the ratio $M_G / M_I$ corresponds to the experimental observation of the universality of the acceleration of free fall for all test masses placed at the same location, which constitutes the basis of the so-called weak equivalence principle (WEP). The equivalence of acceleration and gravitation and the apparent elimination of gravity in mechanical experiments with particles in free fall asserted by WEP amounts to nothing more than the equality of gravitational and inertial masses. The experimental evidence for this equality is excellent. According to modern versions of the Eötvös experiment, deviations from $M_G / M_I = 1$ are smaller than a few parts in $10^{13}$, and planned satellite tests (STEP) are expected to explore this to 1 part in $10^{18}$ [Will, 2005].

In contrast, the experimental evidence for the so-called strong equivalence principle (SEP) is spotty. Simply stated, SEP asserts that in a sufficiently small reference frame in free fall, all gravitational effects are eliminated, in mechanical and any other experiments. However, even in a small reference frame, sufficiently precise test will reveal the curvature of spacetime by the presence of tidal forces and tidal torques [Misner et al., 1973, Section 16.5; Ohanian, 1977] or by directly measurable deviations from flat spacetime geometry (measurable in principle, if not yet in practice). This means that SEP is valid only if we place artificial, statutory limits on the size of the reference frame or on the precision of our experiments.[1] Just a few years after Einstein formulated General Relativity, Eddington gave a dismissive opinion about SEP:

> It is essentially a hypothesis to be tested by experiment as opportunity offers. Moreover, it is to be regarded as a suggestion, rather than a dogma admitting no exceptions. It is likely that some of the phenomena will be determined by comparatively simple equations in which the components of the curvature of the world will not appear…But there are more complex phenomena governed by equations in which the curvatures of the world are involved; terms containing these curvatures will vanish in the equations summarising experiments made in a flat region, and would have to be reinstated in passing to the general equations. Clearly there must be some phenomena of this kind which discriminate between a flat world and a curved world; otherwise we could have no knowledge of world curvature. For these the Principle of Equivalence breaks down…The Principle of Equivalence offers a suggestion for trial, which may be expected to succeed sometimes, and fail sometimes. [Eddington, 1924, pp. 40, 41]

In short, SEP is merely an *Ansatz*, for the use of minimal couplings whenever possible. But sometimes nonminimal couplings—that is, direct couplings between matter and the Riemann tensor—cannot be avoided, as in the case of the tides or the case of spin precession of nonspherical bodies caused by tidal torques. And sometimes nonminimal couplings are desirable for fundamental theoretical reasons, such as the modification of the energy-momentum tensor of matter proposed by Callan, Coleman, and Jackiw [1970] or the direct coupling of the scalar field to the curvature employed in the Brans-Dicke theory [1961]. Nonminimal couplings always violate SEP, but they do not necessarily violate WEP, although they often lead to a partial violation of WEP, that is, a violation for bodies containing appreciable amounts of gravitational self-energy, but not for bodies of small mass ("test masses") with insignificant amounts of gravitational self-energy.

A distinction is occasionally made between the active gravitational mass (the source of the force) and the passive gravitational mass (the recipient of the force), so the gravitational force exerted by a pointlike mass $M_A$ on a pointlike mass $m_P$ can be written as $GM_A m_P / r^2$. But this distinction between active and passive mass is gratuitous, because Newton's third law, or momentum conservation, demands that $M_A m_P = M_P m_A$, which implies that the ratio of active to passive gravitational mass is a universal constant, $M_A / M_P = m_A / m_P = \beta$. The magnitude of the equal gravitational action and reaction forces can then be written as $(G / \beta) M_A m_A / r^2$; and, with a redefinition of $G$, this simply reduces to the usual expression $GM_G m_G / r^2$, with a gravitational mass $M_G \equiv M_A = M_P$ (and likewise for $m_G$).

In a Lagrangian-based theory with no external forces, momentum is necessarily conserved, and Newton's third law is valid, so the active and passive gravitational masses necessarily are the same. However, there can be apparent small violations of Newton's third law caused by retardation effects, gravitomagnetic forces, and radiation-reaction forces, which transfer momentum to the gravitational fields and create the illusion of a loss of momentum from the system. These effects are extremely small; for instance, for two masses $M$ and $m$ moving at right angles at speeds $V$ and $v$, the mutual gravitomagnetic forces predicted by General Relativity differ by about $(GMm / r^2)$ $\times (Vv / c^2)$, which is smaller by a factor $Vv / c^2$ than the ordinary gravitational force. Experimental and observational attempts to find evidence for an inequality of active and passive mass or a violation of momentum conservation were not designed to detect these small, predicted effects; they were based on a vague hope for other, larger effects arising from some new and different mechanism (for instance, some alternative theories of gravitation, not based on Lagrangians, produce violations of momentum conservation). These attempts failed to find anything of interest, to within the limits of their experimental and observational errors [Kreuzer, 1968; Bartlett and Van Buren, 1986].

The universality of free fall motivates the geometric interpretation of gravity, with small test masses moving along geodesics of a curved spacetime geometry. In General Relativity, this geodesic motion can be shown to be a consequence of the "conservation" law $T_\mu{}^\nu{}_{;\nu} = 0$ for nongravitational matter. But this is really a *nonconservation* law—it reveals to what extent the energy-momentum of the nongravitational matter is *not* conserved. Written out in full, it becomes

$$\frac{\partial}{\partial x^\mu} T_\nu{}^\mu = \Gamma^\alpha_{\nu\mu} T_\alpha{}^\mu - \Gamma^\mu_{\alpha\mu} T_\nu{}^\alpha \tag{1}$$

which determines the rate at which the nongravitational matter receives energy and momentum from the gravitational field (the equation is analogous to the equation for the rate of change of the momentum of a particle, $dp/dt = F$).[2] Most writers on General Relativity fail to acknowledge that Eq. (1) is not so much a conservation law, as a law for energy transfer; Weinberg [(1972), p. 166] and Padmanabhan [2010, p. 213] are commendable exceptions. Only in a local geodesic coordinate frame (that is, in freely falling coordinates with $\Gamma^\alpha_{\mu\nu} = 0$) does Eq. (1) reduce to the standard form of a conservation law, $\partial T_\nu{}^\mu / \partial x^\mu = 0$, which shows that the gravitational field delivers no energy or momentum to the nongravitational matter. From this we immediately recognize that the energy and momentum of a small test mass (for which energy contributions of order $M^2$ can be neglected) are constant, so the body remains at rest in the freely falling coordinates and therefore moves along a geodesic.

The implicit assumption in this way of establishing geodesic motion is that the gravitational field of the test mass itself is weak enough so it can be treated as a linear perturbation in the external field. If both the self field and the external field are weak, as in the case of orbiting bodies in the Solar System, then this linearity is of course well satisfied. Linearity allows a sharp and clean separation between the external field and the self field, and it allows an unambiguous identification of the freely falling reference frame in the external field. When linear superposition fails (e.g., two neutron stars in close orbits), geodesic motion also fails.[3]

The linear approximation to General Relativity reproduces the orbital motion of particles (and, more generally, of spherical bodies) found in Newtonian gravitation, and the linear approximation also yields the (non-Newtonian) deflection and retardation of light rays. However, we need to go beyond the linear approximation to extract the perihelion precession, and we also need to go beyond the linear approximation to account for the observational fact that gravitational energy gravitates, that is, the gravitational energy of a body contributes to the gravitational and the inertial masses in the same way. This is an issue that lies outside the realm of Newton's gravitational theory, but high-precision measurements of the Earth-Moon distance by laser ranging have shown that the (negative) gravitational self-energies of the Earth and the Moon do not affect their relative acceleration. This indicates that in these bodies the gravitational self-energy decreases $M_G$ and $M_I$ by the same amount, so as to preserve the equality $M_G = M_I$. The observational data from lunar laser ranging establish that any inequality between $M_G$ and $M_I$ arising from the macroscopic gravitational self-energy $\Omega$ of the Earth is no larger than $4 \times 10^{-4} |\Omega|$. Furthermore, observational data from orbital motions of several pulsar-white-dwarf binary systems establish that any inequality between $M_G$ and $M_I$ arising from the gravitational self-energies of neutrons stars is no larger than $5 \times 10^{-3} |\Omega|$ [Will, 2005, Section 3.6.1].

In the theory of General Relativity we can prove the exact equality of the gravitational and the inertial masses. This theorem applies to the *active* gravitational mass, that is, the gravitational mass as a source of gravitational fields. In General Relativity and in other geometric theories of gravity, the gravitational mass must always be taken to be the *active* gravitational mass, because in such theories there is no passive gravitational mass. A body in a gravitational field moves in response to the curved spacetime geometry, not in response to an impressed gravitational force; thus, the mass of the body is not a receptor of gravitational force, and passive mass is a meaningless concept—it is merely an artifact of the Newtonian approximation. However, when General Relativity gives us a value for the active gravitational mass, we can assume—by momentum conservation—that the Newtonian passive gravitational mass coincides with this active mass. Thus, the exact result $M_G = M_I$ for the active gravitational mass also applies, within the Newtonian approximation, to the passive gravitational mass.

The first theoretical proof of the equality $M_G = M_I$ was given by Eddington [1924, Section 57], but only for the case of a linear approximation for the gravitational field equations (weak-field approximation, and no gravitational self-energy). A more general proof, with the full nonlinear Einstein equations, was subsequently given by Tolman [1934, Section 91] and this proof was later reproduced, with some changes in notation, by Møller [1952, p. 343] and by Zatzkis [1951]. However, all these early attempts at establishing $M_G = M_I$ were incomplete, because they took for granted that the tensor appearing on the right side of the Einstein equations, as source of gravitational fields, is indeed the physical energy-momentum tensor. In the Lagrangian formulation of the Einstein field equations, the source tensor is $T^{\mu\nu} \equiv -(2/\sqrt{-g})\delta\mathcal{L}_m / \delta g_{\mu\nu}$, where $\mathcal{L}_m$ is the Langrangian density for nongravitational matter, and it is not self-evident that this variational derivative equals the canonical energy-momentum tensor, whose volume integral is known to yield the energy-momentum four-vector for the system.

For any given physical system—such as electro-magnetic fields, or classical point particles, or a classical ideal fluid—it is, of course, possible to verify the concordance of the source tensor and the canonical energy-momentum tensor by explicit calculation, and presumably Tolman and Møller had this in mind when they interpreted the volume integral of $T_0^{\ 0}$ as the inertial mass.[4] But this is an unsatisfactory approach, because it hinges on knowing the full internal dynamics (that is, the full Lagrangian) for the gravitating system. Even today, we have no certain knowledge of the full Lagrangian for subatomic matter. We could perhaps accept the Lagrangian of the Standard Model of particle physics (with its multitude of quarks and gauge bosons and their messy interactions) as an *Ansatz*, and we could then verify the concordance of the source tensor and the canonical energy-momentum tensor by explicit calculation. But much of the effort in theoretical physics in the last thirty has been devoted to finding a more fundamental description of elementary particles and their dynamics, and few physicists believe that the Standard Model is the final word. For a general, complete proof of $M_G = M_I$ it is better to establish the concordance of the source tensor and the canonical energy-momentum tensor by general arguments, valid for any (reasonable) Lagrangian, rather than by tedious case-by-case calculations.

I first published a complete general proof of $M_G = M_I$ in a paper [Ohanian, 1973] on the gravitational properties of the "new-improved" energy-momentum tensor proposed by Callan, Coleman, and Jackiw [1970]. Although that paper focused on the case of a scalar field, the proof actually did not make use of the details of the Langrangian; instead, it relied on the general invariance of the Lagrangian and Noether's theorem, so it is quite general and is also valid for nonscalar fields. This proof is here reproduced in Sections II, III, and Appendix 1, where it is extended to Lagrangians with arbitrary nonlinear combinations of second derivatives. My proof does not include classical particles explicitly, because they can be regarded as limiting cases of quantum particles, and the latter can be described by wave fields. Hence all forms of matter in the real world can be described by fields, and in Appendix 1 all matter is treated realistically as interacting classical or quantum fields, without restricting the number or character of such fields in any way whatsoever, except that each field is assumed to belong to linear representation of the Lorentz group (that is, each field is a scalar, vector, tensor, or spinor field some rank or another, or a superposition of those).

The only significant restriction is that there must be no extra, long-range gravitational fields—such as the Brans-Dicke scalar field—that modify the gravitational interactions of Einstein's theory and involve a nonminimal coupling to the curvature tensor. Long-range fields, such as electromagnetic fields, with indirect gravitational effects via the gravitational attraction of the field energy, are of course permitted; and so are nonminimal couplings by short-range fields, that is, a range much shorter than the distance at which the asymptotic gravitational field is measured. This means that, in General Relativity, an explicit violation of the strong principle of equivalence by short-range fields with nonminimal couplings does not undermine the weak principle of equivalence.

Soper [1976, Section 12.5] later gave a similar proof, but his proof is valid only for the limiting case of flat spacetime, $g_{\mu\nu} \to \eta_{\mu\nu}$; and, in contrast to my proof, it does not consider the possible presence of direct nonminimal couplings with second-order derivatives in the Lagrangian, which lead to third-order derivatives in the conservation law for the energy-momentum tensor. More recently, Gamboa-Saravi [2004] also gave a similar proof, free of the restriction $g_{\mu\nu} \to \eta_{\mu\nu}$, but still without nonminimal couplings. Finally, Zhang [2005] gave yet another similar proof that included spinor fields,[5] but, again, without nonminimal couplings.

One good reason for the exploration of nonminimal couplings in General Relativity is that this lays the groundwork for the exploration of nonminimal couplings in alternative theories of gravitation. In scalar and vector theories of gravitation, these couplings give rise to interesting violations of the equality of gravitational and inertial masses, and it is therefore of special interest to consider the implications of such couplings in General Relativity.

Section IV deals with some special examples: the case of a static mass distribution (already treated by Tolman, long ago), and the quite curious case of the maximal Schwarzschild solution, or wormhole, which apparently has an inertial mass twice as large as the gravitational mass.

Section V re-examines the relationship between the gravitational and the inertial masses in the Brans-Dicke theory and shows that these masses differ by a term involving the gravitational self-energy. This difference is called the Nordtvedt effect, but we will see that the original calculations by Nordtvedt, and all the other calculations that followed it, were wrong in detail [Nordvedt, 1968; Will, 1993, Section 6.2]. These calculations yielded a difference proportional to the gravitational self-energy, but with a wrong coefficient of proportionality. The mistake arose from a careless neglect of the scalar-field contribution to the energy-momentum tensor and also a failure to include in the PPN treatment of the equation of motion the energy and momentum transfers between the external gravitational field and the internal gravitational and scalar fields of the system.

Available experimental and observational constraints compel the Brans-Dicke theory into close concordance with General Relativity, and this has made the theory uninteresting. However, the mistakes in the calculation of gravitational and inertial masses in the Brans-Dicke theory are typical of many other alternative theories of gravitation, and for that reason the Brans-Dicke theory provides an instructive lesson about what *not* to do when dealing with energy-momentum tensors. Section V ends with a brief discussion of the application of our results to vector-tensor and to scalar-vector-tensor theories.

Given the persistent mistakes in treatments of the energy-momentum tensor in General Relativity and in alternative theories of gravitation, I deliberately adopt a frankly pedagogical style in this paper, with a full presentation of all details. I hope this will expose all mistakes (including my mistakes, if any).

## II. INERTIAL MASS IN GENERAL RELATIVITY WITH NONMINIMAL COUPLING

General Relativity is based on two fundamental assumptions: 1) All gravitational effects arise in their entirety from a dynamic curved spacetime geometry and 2) all the fields that appear in the theory are dynamic (no fixed background geometry or any other fixed field) with field equations that possess invariance under general coordinate transformations (that is, the manifold mapping group is a symmetry group of the theory). Accordingly, the action integral for the field equations must contain only one gravitational field (the tensor field $g_{\mu\nu}$ and no others), and it must be a scalar under general coordinate transformations.

We can write the action integral as

$$\int \left( \frac{1}{16\pi G} \sqrt{-g}\, R + \mathcal{L}_{\min} + \sqrt{-g}\, f \right) d^4 x \tag{2}$$

where $\mathcal{L}_{\min}$ is the usual minimally-coupled Lagrangian density for the matter fields, constructed from the special-relativistic matter Lagrangian by the familiar "comma to semi-colon" rule, and $f$ is a scalar function that describes extra nonminimal couplings between the matter fields and second derivatives of the metric tensor. We can also include in $f$ extra terms involving nonlinear combinations of second derivatives of the metric tensor, with or without matter fields, that represent more drastic deviations from Einstein's equations, for instance perhaps a term proportional to $R^2$. Such nonlinear terms are believed to arise from quantum fluctuations of matter, and they might play a crucial role during the very earliest stages of the Big Bang. For the purpose of the calculations in this section and Appendix 1, it is convenient to adopt the notation $\mathcal{L}_m \equiv \mathcal{L}_{\min} + \sqrt{-g}\, f$ for the sum of the minimally-coupled Lagrangian density and all the extra nonminimal second-derivative terms. The mathematical formalism permits us to regard this sum as a single term, in which first and second derivatives, linear or nonlinear, can all be treated jointly.

The curvature scalar $R$ multiplied by $\sqrt{-g}$ can be written as a sum of two terms,

$$R\sqrt{-g} = \sqrt{-g}\, g^{\alpha\beta} (\Gamma^{\sigma}_{\alpha\rho}\Gamma^{\rho}_{\beta\sigma} - \Gamma^{\rho}_{\alpha\beta}\Gamma^{\sigma}_{\rho\sigma}) + [g^{\alpha\beta}\sqrt{-g}\, (-\Gamma^{\rho}_{\alpha\beta} + \delta^{\rho}_{\alpha}\Gamma^{\sigma}_{\beta\sigma})]_{,\rho} \equiv \mathcal{L}_G + S^{\rho}{}_{,\rho}$$

Here the first term $\mathcal{L}_G$ is the usual truncated gravitational Lagrangian density, quadratic in the first derivatives of the metric tensor, and the second term is a pure divergence that contains all the second derivatives. The action integral then becomes

$$\int \left( \mathcal{L}_G + S^{\rho}{}_{,\rho} + \mathcal{L}_m \right) d^4 x \tag{3}$$

Although the divergence $S^{\rho}{}_{,\rho}$ does not contribute anything to the field equation, we cannot discard it because it plays an essential role in preserving the scalar characteristics of the action integral, which will be important for our examination of the energy-momentum tensor. However, $S^\rho{}_{,\rho}$ consists of second derivatives of the same type as other second derivatives in $\mathcal{L}_m$, and we can therefore include it in $\mathcal{L}_m$, so it disappears from sight and we can ignore it until Appendix 1.

The action integral (3) leads to the Einstein equations for the gravitational field:

$$-\frac{1}{8\pi G}\left(R^{\mu\nu}-\tfrac{1}{2}g^{\mu\nu}R\right)=-\frac{2}{\sqrt{-g}}\frac{\delta\mathcal{L}_m}{\delta g_{\mu\nu}} \tag{4}$$

or

$$-\frac{1}{8\pi G}\left(R^{\mu\nu}-\tfrac{1}{2}g^{\mu\nu}R\right)=T^{\mu\nu} \tag{5}$$

Here the tensor

$$T^{\mu\nu}\equiv-\frac{2}{\sqrt{-g}}\frac{\delta\mathcal{L}_m}{\delta g_{\mu\nu}} \tag{6}$$

is usually called the energy-momentum tensor of matter.

Note that in Eq. (6), the variations in the 10 components of the symmetric tensor $g_{\mu\nu}$ must be considered as independent; thus, this variational derivative has no meaning in a spacetime with a fixed, prescribed geometry (e.g., flat geometry), in which the only possible variations of the metric tensor are those that can be generated by coordinate transformations. In other words, Eq. (6) makes sense only as part of a theory with a dynamic geometry. Equation (6) is also valid for spinor fields, but the variational derivative $\delta\mathcal{L}_m/\delta g_{\mu\nu}$ must then be expressed in terms of vierbein fields. This makes things a bit more messy, but poses no problems of principle, and does not change any of the following results.

Since $T^{\mu\nu}$ is merely a variational derivative, it is not immediately obvious that it has anything to do with energy-momentum, and it is therefore premature to call it energy-momentum tensor. In Eq. (5), $T^{\mu\nu}$ plays the role of a source for the gravitational field, and we will therefore provisionally call it the gravitational source tensor. For the proof of $M_G=M_I$, we need to establish the relationship between this source tensor $T^{\mu\nu}$ and the actual, physical, energy-momentum tensor. We will see that the integral of the energy density $T_0{}^0$ associated with the source tensor supplemented by the energy density of the gravitational field equals the actual, physical energy derived from the canonical energy momentum tensor. This means that the source tensor does indeed give us the physical energy-momentum density, except perhaps for spurious terms (superpotentials) whose volume integral is zero, so they contribute nothing to the total energy or momentum of the system.

The proof of $M_G=M_I$ involves three steps: 1) show that the volume integral of the source tensor $T_0{}^0$ supplemented by the energy density of the gravitational field can be expressed in terms of the asymptotic components of the metric tensor at large distance [Eqs. (8) - (13)]; 2) show that the volume integral in the first step equals the inertial mass derived from the actual, physical energy-momentum tensor [Eqs. (14) - (18) and Appendix 1]; 3) show that the gravitational mass can be expressed in terms of the asymptotic components of the metric tensor, and compare this asymptotic expression with the expression found in the first step [Eqs. (20) - (22)].

The first two steps in combination correspond to the following theorem:

*For a gravitating body of finite extent at rest in a coordinate frame in an asymptotically flat spacetime, the inertial mass can be expressed as an integral over a large, distant surface enclosing the body,*

$$M_I = \frac{1}{16\pi G} \int \left( \frac{\partial h_{nn}}{\partial x^k} - \frac{\partial h_{kn}}{\partial x^n} \right) dS_k \tag{7}$$

*where the gravitational fields $h_{kn}$ (for $k, n = 1, 2, 3$) represent the asymptotic deviations of order $1/r$ from flat spacetime expressed in rectangular coordinates, with a metric tensor $g_{kn} \rightarrow -1 + h_{kn}$.*

We will see that this theorem is a consequence of Einstein's equations and coordinate invariance symmetry of the action integral. Equation (7) can be regarded as Gauss' law for the gravitational field, because it is analogous to Gauss' law for the electric field. Pauli called it the "flux theorem." Note, however, that the surface integral in Eq. (7) gives us the *inertial mass*, whereas by analogy with the electric case we might have expected that it should give the gravitational mass (that is, the "gravitational charge"). Of course, in General Relativity, we will ultimately find that the inertial mass equals the gravitational mass, so all's well that ends well.

Equation (7) is sometimes called the ADM mass formula, and it has been attributed to Arnowitt, Deser, and Misner [1962]. Versions of this formula were already known earlier; with somewhat different notation it appears in publications by Møller [1952, p. 343], Zatzkis [1951], and Goldberg [1958]. However, as already mentioned in Section I, all these publications gave only incomplete proofs—they all took it for granted that the volume integral of the $_0{}^0$ component of the gravitational source tensor agrees with the inertial mass. They made no attempt to demonstrate this concordance, which is by far the most difficult step of the proof.

Arnowitt et al. demonstrated this concordance, but only for the case of a body consisting of pointlike classical particles interacting with each other by electromagnetic fields. Their proof is somewhat opaque, because it was merely a by-product of a complicated reformulation of General Relativity in terms of a set of four canonical variables that serve as replacements of the usual description of gravitation in terms of the components of the metric tensor $g_{\mu\nu}$. Furthermore, their proof applies only to General Relativity, with purely minimal coupling (no direct coupling to the Riemann tensor), so it does not lend itself to an exploration of why and how the equality of inertial and gravitational mass fails in the Brans-Dicke and other alternative theories of gravitation. Møller did not restrict his energy-momentum tensor to any specific kind of material system, but he assumed, without proof, that the source tensor for the Einstein field equations coincides with the physical energy-momentum tensor, and he does not even relate this tensor to a Lagrangian.

To prove the gravitational Gauss' law (7), we want to express Einstein's field equation in an integral form. For this purpose, we move some terms from the left side to the right side of the field equation, so Eq. (5) becomes [6]

$$\frac{1}{16\pi G}\frac{\partial}{\partial x^\beta}\left[\frac{g_{\mu\sigma}}{\sqrt{-g}}\frac{\partial}{\partial x^\alpha}(-g)\left(g^{\nu\sigma}g^{\alpha\beta}-g^{\nu\alpha}g^{\sigma\beta}\right)\right]=\sqrt{-g}\,t_\mu^{\ \nu}+\sqrt{-g}\,T_\mu^{\ \nu} \tag{8}$$

Here $\sqrt{-g}\,t_\mu^{\ \nu}$ is the canonical energy-momentum tensor for the gravitational field,

$$\sqrt{-g}\,t_\mu^{\ \nu}=\frac{1}{16\pi G}\left(g_{\alpha\beta,\mu}\frac{\partial \mathcal{L}_G}{\partial g_{\alpha\beta,\nu}}-\delta_\mu^{\ \nu}\mathcal{L}_G\right) \tag{9}$$

This is usually called the Einstein pseudo-energy-momentum tensor density. It is a sum of terms quadratic in the first derivatives of $g_{\alpha\beta}$; a convenient explicit expression for $t_\mu^{\ \nu}$ is given by Goldberg [1958, Eq. (3.5)]. Note that the Einstein pseudo-tensor is not symmetric, and it differs from the Landau-Lifshitz energy-momentum tensor [Landau and Lifshitz, 1962, p. 343]. The latter is symmetric but it is of questionable physical significance, because its volume integral does not lead to a four-component energy-momentum quantity that transforms as a vector, but as a vector density. Several alternative symmetric energy-momentum tensors can be constructed, for the formulation of the angular-momentum conservation law [Goldberg, 1958].

The $_0^0$ component of Eq. (8) is

$$\frac{1}{16\pi G}\frac{\partial}{\partial x^\beta}\left[\frac{g_{0\sigma}}{\sqrt{-g}}\frac{\partial}{\partial x^\alpha}(-g)\left(g^{0\sigma}g^{\alpha\beta}-g^{0\alpha}g^{\sigma\beta}\right)\right]=\sqrt{-g}\,t_0^{\ 0}+\sqrt{-g}\,T_0^{\ 0} \tag{10}$$

In this equation, the terms with $\beta=0$ cancel out in the summation over $\beta$, so the equation reduces to

$$\frac{1}{16\pi G}\frac{\partial}{\partial x^k}\left[\frac{g_{0\sigma}}{\sqrt{-g}}\frac{\partial}{\partial x^\alpha}(-g)\left(g^{0\sigma}g^{\alpha k}-g^{0\alpha}g^{\sigma k}\right)\right]=\sqrt{-g}\,t_0^{\ 0}+\sqrt{-g}\,T_0^{\ 0} \tag{11}$$

Here the left side is a 3-D divergence, similar to the 3-D divergence in the differential form of Gauss' law for the electric field, $\partial E^k/\partial x^k=4\pi\rho$. As in the electric case, we can transform the differential equation (11) into an integral equation by integrating over a 3-D volume enclosed in a surface S. The left side then becomes a surface integral,

$$\frac{1}{16\pi G}\int \frac{g_{0\sigma}}{\sqrt{-g}}\frac{\partial}{\partial x^{\alpha}}(-g)\left(g^{0\sigma}g^{\alpha k}-g^{0\alpha}g^{\sigma k}\right)dS_{k}=\int \left(\sqrt{-g}\,t_{0}^{\ 0}+\sqrt{-g}T_{0}^{\ 0}\right)d^{3}x \qquad (12)$$

The factor of $\sqrt{-g}$ in the volume integrals on the right side seems odd here—for a 3-D integration, we expect a factor of $\sqrt{^{(3)}g}$ to convert the coordinate volume element into a metric volume element. However, in the linear, Newtonian approximation, it is easy to check that the factor of $\sqrt{-g}$ is correct, because in this approximation, the metric tensor is diagonal and $\sqrt{-g}=\sqrt{-g_{00}g_{11}g_{22}g_{33}}=\sqrt{(1+2\Phi)}\sqrt{^{(3)}g}\cong(1+\Phi)\sqrt{^{(3)}g}$, so the second term in the integrand includes a contribution $\Phi T_{0}^{\ 0}$, that is, the energy density (or mass density) multiplied by the Newtonian gravitational potential $\Phi$, which equals the gravitational potential energy density, exactly as expected. Apart from this (negative) gravitational potential energy, there is also (positive) gravitational field energy, included in the first term in the integrand on the right side of Eq. (12). In the Newtonian approximation, this gravitational field energy density equals $(\nabla\Phi)^{2}/8\pi G$. The net gravitational energy contribution to the right side of Eq. (12) is negative.

Now suppose that the volume comprises the entire gravitating system and a large surrounding volume whose boundary surface $S$ is far away, so the linear approximation for the gravitational field is applicable there. If we select coordinates in which the gravitating system is at rest, we can approximate $g_{\mu\nu}\cong\eta_{\mu\nu}+h_{\mu\nu}$, where the diagonal components are of order $1/r$, and the off-diagonal components of $g_{\mu\nu}$ are zero, except perhaps for negligible terms of order $1/r^{2}$ (arising from rotational angular momentum or velocities within the system or from a quadrupole moment). With $g_{\mu\nu}=\eta_{\mu\nu}+h_{\mu\nu}$, $g^{\mu\nu}\cong\eta_{\mu\nu}-h_{\mu\nu}$, and $-g\cong1+h_{00}-h_{nn}$, Eq. (12) then becomes

$$\frac{1}{16\pi G}\int \left(\frac{\partial h_{nn}}{\partial x^{k}}-\frac{\partial h_{nk}}{\partial x^{n}}\right)dS_{k}=\int \left(\sqrt{-g}\,t_{0}^{\ 0}+\sqrt{-g}T_{0}^{\ 0}\right)d^{3}x \qquad (13)$$

It remains to establish that the integral on the right side of Eq. (13) equals the inertial mass. By definition, the inertial mass is the energy in the rest frame (or the zero-momentum frame) of the system. In any system with equations of motion derivable from a Lagrangian, the energy is unambiguously defined in terms of the Lagrangian density $\mathcal{L}=\mathcal{L}_{G}+\mathcal{L}_{\mathrm{m}}$ by a formula similar to that for the Hamiltonian for a system of particles:

$$M_{I}=\int \left[Q_{i,0}\frac{\partial}{\partial Q_{i,0}}(\mathcal{L}_{G}+\mathcal{L}_{\mathrm{m}})-(\mathcal{L}_{G}+\mathcal{L}_{\mathrm{m}})\right]d^{3}x \qquad (14)$$

Here the quantities $Q_{i}$, $i=1,2,3,...,$ represent the components of the fields, including the components of $g_{\mu\nu}$ and the components of all the classical and quantum fields that occur in the gravitating system. The integrand on the right side of Eq. (14) is the $_0{}^0$ component of the canonical energy-momentum tensor density $\mathcal{T}_\mu{}^\nu$, calculated from the entire Lagrangian, that is, including both matter and gravitation.[7] However, for the sake of simplicity, let us pretend that the Lagrangian $\mathcal{L}_m$ in Eqs. (14), (15) and (18) contains only first derivatives; the more complicated expression for the energy-momentum tensor that includes second derivatives is given in Appendix 1. Under this pretense,

$$\mathcal{T}_\mu{}^\nu = Q_{i,\mu} \frac{\partial}{\partial Q_{i,\nu}}(\mathcal{L}_G + \mathcal{L}_m) - \delta_\mu^\nu(\mathcal{L}_G + \mathcal{L}_m) \tag{15}$$

This canonical energy-momentum tensor obeys the standard conservation law

$$\partial_\nu \mathcal{T}_\mu{}^\nu = 0 \tag{16}$$

In terms of $\mathcal{T}_0{}^0$, Eq. (14) takes the concise form

$$M_I = \int \mathcal{T}_0{}^0 d^3x \tag{17}$$

The identification of this integral as energy is confirmed by the interchangeability of the "amounts of $\int \mathcal{T}_0{}^0 d^3x$" of the field system with the amount of energy of any classical particles that interact with this system. For instance, when a particle exchanges some amount of $\int \mathcal{T}_0{}^0 d^3x$ with the system in an inelastic scattering process, the conservation law obeyed by the canonical energy-momentum tensor during the interaction with the particle ensures that this amount corresponds to the change of kinetic energy of the outgoing vs. the incoming particle.[8]

Comparing the right sides of Eqs. (13) and (14), we see that the contributions from $\mathcal{L}_G$ coincide, because $\sqrt{-g}\, t_\mu{}^\nu$ equals the canonical energy-momentum tensor density [see Eq. (9)]. It therefore only remains to establish that the contributions from $\mathcal{L}_m$ also coincide, that is,

$$\int \left( -2g_{\alpha 0} \frac{\delta \mathcal{L}_m}{\delta g_{\alpha 0}} \right) d^3x = \int \left( Q_{l,0} \frac{\partial \mathcal{L}_m}{\partial Q_{l,0}} - \mathcal{L}_m \right) d^3x \tag{18}$$

A general proof of this equality is given in Appendix 1, for an arbitrary combination of fields with arbitrary interactions, subject to the restriction that all the field equations are derivable from a Lagrangian (a restriction obeyed by all the theories of particles and interactions we know of). The proof given in Appendix 1 includes a generalization to Lagrangians with second-order derivatives of the fields, such as the second-order derivatives of the metric tensor typical of nonminimal couplings between matter and gravitation [see Eq. (A1) for the generalized expression for the canonical tensor including second derivatives]. The essential ingredient in this proof is the invariance of the Lagrangian under general coordinate transformations (without such invariance, it seems impossible to construct a general proof; this is a serious and apparently insuperable defect of those alternative theories of gravitation that lack this invariance).

In consequence of the equality (18), the right side of Eq. (13) then equals $M_I$, which completes the proof of the theorem.

The right side of Eq. (7) can be expressed in terms of the Killing vector $\zeta^\alpha = (1, 0, 0, 0)$, $\zeta_\alpha = (g_{00}, 0, 0, 0)$ that characterizes the invariance of the asymptotic spacetime under time translation. This yields an elegant covariant expression for the inertial mass,[9]

$$M_I = \frac{1}{8\pi G} \int (\zeta^\alpha N^\beta - \zeta^\beta N^\alpha) \zeta_{\alpha;\beta} dS \qquad (19)$$

where $N^\alpha$ is the outward pointing unit vector normal to the spatial surface element $dS$. If the spacetime is not only asymptotically static, but static throughout the entire volume, then the Killing vector exists everywhere and it is possible to generalize Eq. (19) so it applies to any arbitrary (not necessarily asymptotically distant) surface enclosing the gravitational sources (see Wald, pp. 288 - 290). However, this is of limited interest, because realistic astrophysical systems are hardly ever exactly static when viewed at short distances. For instance, the gravitational field of the Solar System can be treated as static at a large distance, but the planetary motions make the gravitational field at shorter distances time-dependent; likewise, the gravitational field of a pulsar can be treated as static at a large distance, but its off-axis rotating magnetic fields imply the presence of time-dependent gravitational fields at short distances.

## III. GRAVITATIONAL MASS

As stated in the Introduction, in General Relativity, the gravitational mass must be taken to be the *active* gravitational mass, because in a geometric theory of gravity, the passive gravitational mass has no exact meaning. When General Relativity gives us a value for the active gravitational mass, we can assume that within the Newtonian approximation this active mass coincides with the Newtonian passive gravitational mass; but this gives only an approximate meaning to the passive mass. To formulate an exact theorem about the gravitational vs. the inertial mass, we will rely exclusively on the active gravitational mass.

The active gravitational mass is simply related to the asymptotic behavior of $g_{00}$, because this completely determines orbital motions in the limit of large distance and low speeds, that is, the Newtonian limit. The asymptotic field $h_{00}$ is proportional to the Newtonian potential,

$$h_{00} = 2\Phi = -\frac{2GM_G}{r} \qquad (20)$$

or, if factors of $c$ are included, $h_{00} = -2GM_G / rc^2$. This can be compared with the solution of the linearized Einstein vacuum field equations, which are valid in the asymptotic region and give us a spherically symmetric field that depends on only one adjustable constant,

$$h_{00} = \frac{A}{r}, \quad h_{kn} = \delta_k^n \frac{A}{r} \tag{21}$$

Accordingly, from Eq. (20), $A = -2GM_G$.

With the value of $A$ fixed in this way, we can then immediately evaluate the inertial mass from the gravitational Gauss law. Under transformations of the spatial coordinates, the integrand of the surface integral in Eq. (7) is a scalar, and it can therefore be evaluated in any asymptotically rectangular coordinate system (subject to the restriction that $g_{kn} = -\delta_k^n +$ small correction ). Equation (21) then leads to the expected final result for the equality of the gravitational and the inertial masses,

$$M_I = \frac{1}{16\pi G} \int \left( \frac{\partial h_{nn}}{\partial x^k} - \frac{\partial h_{kn}}{\partial x^n} \right) dS_k = \frac{1}{16\pi G} \int \left( 3A \frac{\partial}{\partial x^k} \frac{1}{r} - A \frac{\partial}{\partial x^k} \frac{1}{r} \right) dS_k = -\frac{A}{2G} = M_G \tag{22}$$

This exact equality of gravitational and inertial masses is a distinctive feature of Einstein's General Relativity, and, in consequence of the results of Appendix 1, it remains valid even in the presence of nonminimal couplings. Note that the proof of this equality hinges on the two basic assumptions of General Relativity: the invariance of the theory under general coordinate transformations (which is crucial for establishing the relation between the source tensor and the canonical energy-momentum tensor; see Appendix 1), and the absence of extra long-range gravitational fields (which would modify the relation (20) between the Newtonian potential and the asymptotic metric tensor). In alternative theories of gravitation, one or both of these assumptions are violated, and the consequent inequality between $M_G$ and $M_I$ is typically proportional to amount of gravitational self-energy in the system (see Section V).

## IV. SOME SPECIAL EXAMPLES

As first shown by Tolman [1934, Section 92], if the solution of the Einstein equation is time-independent and free of singularities, the inertial mass (7) (and therefore also the gravitational mass) can be expressed as a volume integral over the diagonal components of the source tensor,

$$M_I = \int (T_0^{\;0} - T_1^{\;1} - T_2^{\;2} - T_3^{\;3}) \sqrt{-g} \, d^3 x \tag{23}$$

This result can be derived by converting volume integral of $T_\mu{}^\mu \sqrt{-g}$ into a surface integral, by exploiting the field equation. In Eq. (23), the volume integral of $T_0{}^0$ represents the energy contribution of matter, and the other terms must therefore represent the contribution from the gravitational self-energy. If the matter is in the form of a fluid characterized by density and pressure, then $T_0{}^0 = \rho$, $T_1{}^1 = T_2{}^2 = T_3{}^3 = -p$, and

$$M_I = \int (\rho + 3p)\sqrt{-g}\, d^3x \qquad (24)$$

In the Newtonian limit, this equation reduces to an integral of the mass density and a (negative) integral that represents the gravitational self-energy. To see this, we approximate

$$(\rho + 3p)\sqrt{-g} \cong (\rho + 3p)\sqrt{(1+2\Phi)}\sqrt{{}^{(3)}g} \cong (\rho + 3p + \rho\Phi)\sqrt{{}^{(3)}g} \qquad (25)$$

The integral of the last term is $2\Omega$, that is, twice the (negative) gravitational self-energy. And by the virial theorem for the pressure of a mass of fluid in equilibrium under the action of its gravity and its pressure, the integral of $3p$ equals $-\Omega$, that is, the negative of the gravitational self-energy. Hence the sum of these two integrals equals $\Omega$, and the net result is as expected,

$$M_I \cong \int \rho \sqrt{{}^{(3)}g}\, d^3x + \Omega \qquad (26)$$

However, the neat formulas (23) and (24) are of limited use. They require a time-independent solution of the Einstein equations, whereas Eq. (7) requires only that the asymptotic $1/r$ part of the metric tensor be time-independent; and they also require knowledge of the metric tensor throughout the matter distribution (this is needed because the solution of the static equilibrium equations for matter involves a concurrent solution of the Einstein equations for the gravitational field, usually done by integration of these differential equations from the center outward). If we have adequate computational techniques to find the metric tensor within the mass distribution, we can just as well continue the computation into the distant asymptotic region, and find the inertial mass by examination of the $1/r$ term in $h_{nk}$; computationally, this is no more difficult than evaluation of the integral (24).[10]

If we want to apply the gravitational Gauss' law to the Schwarzschild empty-space solution, we must be careful to select a volume of integration (that is, a spacelike hypersurface) that avoids the real, physical singularity at $r = 0$. In the case of the maximal Schwarzschild solution—which represents two universes, or maybe two widely separated parts of our own universe, joined by a "wormhole" or a "throat"—it is possible to select a hypersurface such that the integration over the volume of the wormhole encounters no singularity. This hypersurface extends from the asymptotically flat spacetime at one end of the wormhole to the asymptotically flat spacetime at the other end. This is best seen by looking at a Kruskal spacetime diagram for the Schwarzschild solution, with coordinates $v$ and $u$, which shows singularity-free spacelike hypersurfaces $v = $ constant in the range $-1 < v < 1$ (hypersurfaces outside of this range encounter singularities; but, by energy conservation, for such hypersurfaces, the integral of the energy density can be presumed to have the same value as for the singularity-free hypersurfaces).

When we integrate Eq. (12) over the volume of the wormhole and convert the left side into a surface integral [as in Eq. (7)], we have to include two separate surface integrals, one around each end of the wormhole.[11] Thus, the value of $M_I$ will now be twice as large, but the value of $M_G$, extracted from the asymptotic behavior of $h_{00}$, will be the same as before. We therefore find the strange result $M_I = 2M_G$ for the gravitational and inertial masses of a wormhole, that is, the wormhole has an abnormally large inertia compared with its gravitational mass. Intuitively, this makes some sense— the nearby end of a wormhole has "extra baggage" attached at the other, distant, end, so it is not surprising that this wormhole is harder to accelerate than a mass distribution with the same external Schwarzschild field, but no extra attached baggage at any distant end. The equality of $M_I$ and $M_G$ will be restored if we evaluate Eq. (20) from the joint gravitational field of both ends of the wormhole, at a distance $r$ much larger than the distance between the ends. But if the ends of the wormhole are widely separated, an observer placed near one end might be so far from the other end that she is not at all aware of that other end, and this observer will perceive a discrepancy between inertial and gravitational mass.

Observationally, the extra inertia means that a wormhole mouth orbiting a star will move with an abnormally low orbital speed, too low by a factor of $\sqrt{2}$ for a given orbital radius (and the motion will not be geodesic!). In principle, this means we can observationally discriminate between a true wormhole (with perfectly empty space in its interior) and a "frozen star," that is, a mass that has collapsed into its own horizon (with a mass distribution in its interior). The latter will have a normal inertial mass, as can be seen by evaluating the inertial mass *before* the collapse, which must be the same as that after the collapse, by energy conservation.

The same is true for the Kerr-Newman black hole. Again, we can select a spacelike hypersurface free of singularities for our volume integration, avoiding the singularity at $r = 0$ (in the maximal empty space Kerr-Newman solution, the singularity is time-like, but does not come into existence until the solution reaches a late stage[12]). This hypersurface begins on one side of the wormhole and extends to the other side, as in the Schwarzschild case, so, again $M_I = 2M_G$.

In a recent paper, Nieuwenhuizen [2007] claimed that for a Kerr-Newman black hole, the passive gravitational mass equals the inertial mass, but the active gravitational mass is twice as large, that is, $M_I = M_G / 2$, which differs from the above result by a factor of 4. It is immediately clear that something is wrong with this claim, because momentum conservation[13] requires a universal value for the ratio of the active and passive masses (and it is then possible to redefine $G$ so as to make these masses equal; see Section I), whereas Nieuwenhuizen claims that the active and passive masses are unequal for a Kerr-Newman black hole. (Observationally, a Kerr-Newman black hole and a normal star of equal inertial masses, equal passive masses, but unequal active masses orbiting about each other would display an erratic orbital motion, with the center of mass executing wild gyrations.) Nieuwenhuizen's calculation of the inertial mass contains several mistakes: he takes into account the inertial mass contributed by the energy density of the electromagnetic field in the region exterior to the horizon, but omits the contribution from the energy density in the interior region; and he completely omits all contributions from the energy density of the gravitational field.[14]

## V. VIOLATIONS OF $M_I = M_G$ IN ALTERNATIVE GRAVITY THEORIES

Alternative theories of gravitation differ from Einstein's theory in that they include extra gravitational fields (scalar, vector, or tensor fields). Some of these theories include extra non-dynamic, fixed fields, prescribed by the inventor of the theory, which destroy the invariance of the theory under general coordinate transformations. Many of these theories have nonminimal couplings—for instance, the Brans-Dicke theory has a nonminimal coupling of its scalar field to the curvature tensor, and so do several other alternative scalar-vector theories. Nonminimal couplings generate extra gravitational interactions, that is, extra interactions between masses, sometimes rather suprising interactions. For instance, the nonminimal coupling of the Callan-Coleman-Jackiw "new improved" energy-momentum tensor [Callan et al., 1970] leads to a short-range extra gravitational interaction [Ohanian, 1973].

A nonminimally coupled field of short range does not affect the long-range behavior of the geometry, and it need not affect the equality of $M_I$ and $M_G$ (if there are no other troubles). But if the nonminimally coupled field is long-range, it affects the geometry and modifies the values of the fields $h_{00}$ and $h_{kn}$ that characterize the weak-field limit at large distances. As we saw in Sections II and III, $h_{00}$ is proportional to the gravitational mass [Eq. (20)], whereas $h_{kn}$ is proportional to the inertial mass [Eq. (7)]. In alternative theories of gravitation, the factors of proportionality are different from those in General Relativity, and this leads to an inequality of the gravitational and inertial masses.

If the ratio of the factors of proportionality is a universal constant (that is, independent of the internal characteristics of the gravitating body), then it is possible to eliminate the inequality of the gravitational and the inertial masses by a redefinition, or renormalization, of the coupling constant $G$ in the gravitational Gauss Law, while keeping the constant $G$ in Eq. (20) fixed at its experimental value. However, when we consider the nonlinearities in the Brans-Dicke theory for bodies other than test masses, with appreciable gravitational self-energies, the ratio of $h_{00}$ and $h_{kn}$ is not a universal constant, which means that $M_I$ and $M_G$ are not equal, and cannot be made equal by a redefinition of $G$. All the proposed alternative theories of gravity that rely on extra, long-range, scalar fields suffer from this defect, and we should expect all of them to violate the equality of inertial and gravitational mass. Of course, because these theories usually have various adjustable constants, it is sometimes possible to enforce an approximate equality of $M_I$ and $M_G$ by a choice of these constants; but such an ad hoc selection of constants is always an indication of a poorly designed theory.

In this section we will examine examples of scalar, vector, and tensor theories and their violations of the equality of inertial and gravitational mass.

**a) Brans-Dicke and other scalar theories.** [NOTE: *To facilitate comparison of our equations with those of Brans-Dicke, we adopt, in this section only, the signature* $-+++$ *for the metric tensor, and we adopt Latin indices to indicate 0, 1, 2, 3. When a summation is restricted to 1, 2, 3, this will be indicated explicitly, as in Eq. (30)*].
Scalar theories have been summarized by Will [1993, Chapter 5]. They all contain a scalar field $\phi$ coupled nonminimally to the Riemann curvature scalar through a term $\phi R \sqrt{-g} / 16\pi$ in the Lagrangian density. This term can be rewritten as a sum of two pieces, $\phi_0 R \sqrt{-g} / 16\pi + (\phi - \phi_0) R \sqrt{-g} / 16\pi$, with a constant $\phi_0$. The first piece is the usual gravitational Lagrangian density $\mathcal{L}_G$, and the second piece is a nonminimal coupling, as in Eq. (2).

Because these theories are endowed with invariance under general coordinate transformations, the general results in Appendix 1 about the relationship between the canonical energy-momentum tensor and the source tensor remain valid. However, the superpotential $W_k^{[n\,l]}{}_{,l}$ now has a nonzero value for the volume integral of its $k=0$, $n=0$ component. This brings about an inequality between the gravitational and inertial masses.

The gravitational field equation of the Brans-Dicke theory is [Brans and Dicke, 1961]

$$\frac{\phi_0}{8\pi}(R_k^{\ n} - \tfrac{1}{2}\delta_k^n R) = \frac{1}{8\pi}(\phi_0 - \phi)(R_k^{\ n} - \tfrac{1}{2}\delta_k^n R) + T_k^{\ n} + \frac{\omega}{8\pi\phi}(\phi_{,k}\phi^{,n} - \tfrac{1}{2}\delta_k^n\phi_{,l}\phi^{,l})$$
$$+ \frac{1}{8\pi}(\phi_{,k}^{\ ;n} - \delta_k^n\phi_{,l}^{\ ;l}) \tag{27}$$

The source tensor on the right side includes terms with second derivatives, which are not present in the canonical energy-momentum tensor (the Lagrangian for the scalar field does not include second derivatives). Thus, there is a difference involving second derivatives of $g_{kn}$ and of $\phi$ between the source tensor and the canonical tensor. This difference arises from the first and the last terms in the source tensor. According to Appendix 1, the difference between these tensor, multiplied by $\sqrt{-g}$ as per Eq. (8), is the superpotential $V_k^{[n\,l]}{}_{,l}$. This superpotential can be determined by explicit calculation of the canonical tensor, but it is more conveniently determined by its general properties: it must have the same second derivatives (with the same coefficients) as $\sqrt{-g} \times$ source tensor; it must be a tensor density under linear transformations; and it must be a total divergence, antisymmetric in $n, l$. This determines the superpotential as

$$V_k^{[nl]}{}_{,l} = \frac{1}{8\pi} \left\{ (\phi - \phi_0) \frac{g_{ks}}{\sqrt{-g}} \Big[ (-g)(g^{ns} g^{rl} - g^{nr} g^{sl}) \Big]_{,r} \right\}_{,l} + \frac{1}{8\pi} \Big[ \delta_k^{\ l} (\phi'^n \sqrt{-g}) - \delta_k^n (\phi'^l \sqrt{-g}) \Big]_{,l}$$

(28)

Note that the existence of this nonzero superpotential shows explicitly that the source tensor for the gravitational field *is not equal to* the energy-momentum tensor. This is a quite general result: whenever the gravitational theory includes a nonminimal coupling, the source tensor for the gravitational field differs from the canonical energy-momentum tensor, and the usual dogma that "energy-momentum is the source of gravitation" is false. If the nonminimally coupled field is of short range, then this deviation from the dogma of General Relativity is harmless—the gravitational mass still equals the inertial mass, and the nonzero superpotential merely means that the energy and momentum are shifted around within the gravitational and scalar fields, without altering the total energy momentum. But we will see that in the Brans-Dicke theory, this deviation is *not* harmless—not only is the energy-momentum shifted around, but the total energy momentum and the total gravitational mass are altered.

The first term in the Brans-Dicke superpotential (28) is unimportant for the calculation of $M_I$, because the volume integral of the $k = 0$, $l = 0$ component merely generates a surface integral that vanishes for $r \to \infty$. But, for a static field, the integration of the second term contributes

$$-\frac{1}{8\pi} \int (\phi^l \sqrt{-g})_{,l} d^3x = -\frac{1}{8\pi} \int \phi'^l{}_{;l} \sqrt{-g} \, d^3x = -\frac{1}{3+2\omega} \int T \sqrt{-g} \, d^3x$$

(29)

where the last equality comes from the Brans-Dicke field equation for $\phi$ [see Eq. (36)].

Taking into account the extra term (29), we see that the volume integral of the source tensor (with an extra factor $\sqrt{-g}$) differs from the inertial mass, and our previous Eq. (7) now becomes

$$M_I + \frac{1}{2+3\omega} \int T \sqrt{-g} d^3x = -\frac{\phi_0}{16\pi} \int \left( \frac{\partial h_{nn}}{\partial x^k} - \frac{\partial h_{kn}}{\partial x^n} \right) dS_k \quad \text{, where } k, \, n = 1, 2, 3$$

(30)

Brans-Dicke give explicit expressions for $h_{kn}$ and $\phi_0$ calculated from a linear approximation. However, these are based on a solution involving the conservation law $T_k^{\ l}{}_{,l} = 0$ for non-gravitational matter, which implies that in the rest frame of a system in internal equilibrium, all the volume integrals of $T_k^{\ l}$ are zero, with the exception of $T_0^{\ 0}$ (the system has no net stresses and no net momentum, only a net energy).[15] This is a reasonable approximation for a test mass, but not for a more complicated system, because it would exclude the presence of internal gravitational forces and internal gravitational energy, and is therefore inadequate to treat the effects of internal gravitational energy on the inertial and gravitational masses.

Instead of the linear approximation, it is better to use the asymptotic forms of the exact spherically symmetric solution obtained by Brans-Dicke [1961]. With correction of a misprint in the solution,[16] we have

$$g_{00} = -\left(\frac{1-B/r}{1+B/r}\right)^{2/\lambda} \rightarrow -1 + \frac{4B/\lambda}{r} \tag{31}$$

$$g_{11} = g_{22} = g_{33} = (1+B/r)^4 \left(\frac{1-B/r}{1+B/r}\right)^{2-(C+1)/\lambda} \rightarrow 1 + \frac{4(C+1)B/\lambda}{r} \tag{32}$$

$$\phi = \phi_0 \left(\frac{1-B/r}{1+B/r}\right)^{C/\lambda} \rightarrow \phi_0 \left(1 - \frac{2CB/\lambda}{r}\right) \tag{33}$$

This solution is characterized by two independent constants, $B$ and $C$, with $\lambda$ related to those two constants by $\lambda = [(C+1)^2 - C(1-\omega C/2)]^{1/2}$. However, for the asymptotic solution, it is convenient to adopt $B/\lambda$ and $C$ as the two independent constants. The former determines the gravitational mass, and the latter the inertial mass, via the gravitational Gauss law. By inspection of Eq. (31), we see that the gravitational mass is

$$M_G = \frac{2}{G}\frac{B}{\lambda} \tag{34}$$

And by evaluating Eq. (30) with the values of $h_{kn}$ given by Eq. (32), we find that the inertial mass is

$$M_I + \frac{1}{2+3\omega}\int T\sqrt{-g}\,d^3x = 2\phi_0(C+1)\frac{B}{\lambda} \tag{35}$$

Because the constants $B/\lambda$ and $C$ are independent, the gravitational mass is independent of the inertial mass, that is, different physical systems will, in general, have different ratios of inertial to gravitational mass. The difference between the gravitational and the inertial mass—or, what is the same thing, the difference between the asymptotic forms (31) and (32)—arises from the presence of the long-range scalar field $\phi$, which modifies the gravitational fields $g_{kl}$ not only in the near region, but also in the distant, asymptotic, region. If instead of the long-range Brans-Dicke scalar field we were dealing with a short-range Yukawa field (with a reasonably small value of the Yukawa mass parameter), the asymptotic gravitational fields would be those of the Schwarzschild solution, and there would be no difference between the gravitational and inertial mass, that is, the WEP would be satisfied even though the coupling $\phi R$ of the scalar field to the curvature tensor violates the SEP.

The ratio of gravitational and inertial mass in the Brans-Dicke theory is actually related to the trace of the energy-momentum tensor of matter, $T \equiv T_k^{\ k}$. To establish this relationship, we need to construct the explicit solution for $\phi$ in terms of $T$, starting with the field equation for $\phi$:

$$\phi_{,k}{}^{;k} = \frac{8\pi}{3+2\omega}T \qquad (36)$$

or

$$\frac{\partial}{\partial x^k}\left(\sqrt{-g}\,g^{kl}\partial_l\phi\right) = \frac{8\pi}{3+2\omega}T\sqrt{-g} \qquad (37)$$

By integration of this field equation over a spherical volume extending to the (large) radius $r$ and then exploiting Gauss' theorem to convert the left side into a surface integral, we immediately find that the asymptotic value for $\phi$ is

$$\phi = \phi_0 - \frac{2}{3+2\omega}\frac{1}{r}\int T\sqrt{-g}\,d^3x \qquad (38)$$

Comparing this explicit solution with Eq. (33), we see that

$$2\phi_0 C\frac{B}{\lambda} = \frac{2}{3+2\omega}\int T\sqrt{-g}\,d^3x \qquad (39)$$

and eliminating the constants $C$ and $B/\lambda$ among Eqs. (33), (35), and (39), we obtain a general (and exact) relationship between $M_G$ and $M_I$:

$$\phi_0 GM_G + \frac{1}{3+2\omega}\int T\sqrt{-g}\,d^3x = M_I \qquad (40)$$

This agrees with the result I obtained years ago by a different method [Ohanian, 1971; Eq. (30) in that paper coincides with Eq. (40) above, but with a different notation [17]].

When dealing with a "test mass"—that is, a bound system of sufficiently small mass, so internal gravitational forces and gravitational energies and any other effects of order $M^2$ can be neglected—the internal equilibrium of the system demands that the integration over the internal stresses gives zero. Therefore

$$\int T\sqrt{-g}\,d^3x \cong \int T_0^0 d^3x = -M_I \qquad (41)$$

The combination of Eq. (41) and the relation (40) between $M_G$ and $M_I$ implies that for test masses

$$\phi_0 GM_G = \frac{4+2\omega}{3+2\omega}M_I \qquad (42)$$

Thus, the ratio of gravitational and inertial mass is a universal constant for such test masses, and we can define the gravitational constant measured in laboratory experiments so that $M_I = M_G$, which requires the relation originally adopted by Brans Dicke,

$$\phi_0 = \frac{1}{G}\frac{4+2\omega}{3+2\omega} \tag{43}$$

In practice, this relation differs little from $\phi_0 = 1/G$, because observations of radio signals of the Cassini spacecraft imply that $\omega \geq 40,000$ [Will 2005], so the coupling constant $\phi_0$ is very near to $1/G$ (and, likewise, all PPN parameters of the Brans-Dicke theory are very near those of General Relativity). However, similar relations arise in other alternative theories of gravitation, such as the recent attempt at a scalar-vector-tensor theory by Moffat [2006; Moffat and Toth, 2009]; in this theory the coupling constant corresponding to the Brans-Dicke $\omega$ is assumed to be smaller, so the renormalization effects become more prominent.

When dealing with an arbitrary body, not necessarily a test mass, we want to evaluate Eq. (40) to second order in $M$, to see how gravitational self-energy from the gravitational and the scalar fields affects the equality of gravitational and inertial masses. For this, we need a more exact relation between $\int T\sqrt{-g}\,d^3x$ and $M_I$, which we can find by examining the integral of the trace of the canonical energy-momentum tensor. This integral consists of contributions from nongravitational matter, from the scalar field, and from the gravitational field,

$$\int \mathcal{T}d^3x = \int\left[\frac{1}{8\pi}(\phi_0-\phi)(-R)+T+\frac{\omega}{8\pi\phi}(-\phi_{,n}\phi^{,n})+t\right]\sqrt{-g}\,d^3x \quad n=1,2,3 \tag{44}$$

As for a test particle, the integration over the internal stresses gives zero, so the integral on the left side of Eq. (44) reduces to $\int \mathcal{T}_0^{\,0}d^3x = -M_I$. On the right side, it suffices to substitute a first-order approximation for $R$, that is,

$$-\frac{\phi_0}{8\pi}R \cong T-\frac{3}{8\pi}\phi_{,k}^{\;\;;k} \tag{45}$$

obtained from the field equation (27) by deleting all second-order terms on the right side. We can also approximate most of the factors $\sqrt{-g} \cong 1$. With these approximations, we obtain the following expression, correct to second order in $M$,

$$-M_I \cong \int\left[\frac{\phi_0-\phi}{\phi_0}\left(T-\frac{3}{8\pi}\phi_{,n,n}\right)+T\sqrt{-g}+\frac{\omega}{8\pi\phi_0}(-\phi_{,n}\phi_{,n})+t\right]d^3x \tag{46}$$

If $\phi$ is time independent, we can substitute $T \cong [(3+2\omega)/8\pi]\phi_{,n,n}$ with $n = 1, 2, 3$ in the first term in the integrand and then integrate by parts, so

$$-M_I \cong \int \left[ T\sqrt{-g} + \frac{1}{8\pi}\frac{\omega}{\phi_0}\phi_{,n}\phi_{,n} + t \right] d^3x \ , \ \ n = 1, 2, 3 \tag{47}$$

According to Eqs. (37) and (41), to first order in $M$, the scalar field $\phi - \phi_0$ is proportional to the Newtonian gravitational potential $\Phi = -\int (G\rho/r)d^3x$; and, to order $M^2$, the integral involving $\phi_{,n}\phi_{,n}$ is therefore equal to

$$\frac{\omega}{8\pi\phi_0 G^2}\frac{4}{(3+2\omega)^2}\int \Phi_{,n}\Phi_{,n}d^3x = -\frac{4\omega}{\phi_0 G(3+2\omega)^2}\Omega = \frac{4\omega}{(3+2\omega)(4+2\omega)}|\Omega| \tag{48}$$

where $\Omega = -(1/8\pi G)\int \Phi_{,n}\Phi_{,n}d^3x = \frac{1}{2}\int \rho\Phi d^3x$ is the Newtonian gravitational self-energy. The integral of $t = t_0{}^0 + t_1{}^1 + t_2{}^2 + t_3{}^3$ can evaluated by starting with the explicit expression for $t_k{}^n$ in terms of the metric tensor and its derivatives [Goldberg, 1958, Eq. 3.5] and substituting the linear approximation, $g_{00} \cong -1 - 2\Phi$ , $g_{11} \cong 1 - 2\Phi(1+\omega)/(2+\omega)$ , etc.[18] A somewhat tedious calculation gives the following result for the integral of $t$:

$$-\frac{\phi_0}{8\pi}\frac{5+8\omega+2\omega^2}{(2+\omega)^2}\int \Phi_{,n}\Phi_{,n}d^3x = \phi_0 G\frac{5+8\omega+2\omega^2}{(2+\omega)^2}\Omega = -\frac{2(5+8\omega+\omega^2)}{(3+2\omega)(2+\omega)}|\Omega| \tag{49}$$

With these expressions, Eq. (46) becomes

$$-M_I = \int T\sqrt{-g}d^3x + \frac{2\omega - 2(5+8\omega+2\omega^2)}{(3+2\omega)(2+\omega)}|\Omega| \tag{50}$$

Upon elimination of $\int T\sqrt{-g}d^3x$ between Eqs. (40) and (50), we find a final result for the relation between $M_G$ and $M_I$ :

$$M_G \cong M_I - \frac{5+7\omega+2\omega^2}{(3+2\omega)(2+\omega)^2}|\Omega| \tag{51}$$

The inequality of gravitational and inertial masses implies that the system does not move on a geodesic. In essence, the deviation from geodesic motion arises because the system receives energy and momentum not only from the external gravitational field, but also from the external scalar field, which acts on the internal gravitational field and generates an extra energy and momentum transfer, that is, an extra force.

Equation (51) differs from the results obtained by Nordvedt and Will from calculations of the equation of motion of a body moving in an external gravitational field by the PPN approximation scheme. The coefficient of $-|\Omega|$ in Eq. (51) is called the Nordvedt parameter, and the PPN scheme yields a value of $\eta_N = 1/(2+\omega)$ for this parameter, which disagrees with the value obtained in Eq. (51) [Will, 1994, Eq. (6.49)]. Taken at face value, this would suggest that there is a discrepancy between the active gravitational mass (calculated here) and the passive gravitational mass (obtained by Nordvedt and Will). But we know these two gravitational masses must be equal because of momentum conservation, and the real reason for the discrepancy is that the equation of motion constructed in the PPN scheme is infested with various fundamental mistakes, some quite obvious, others somewhat subtle. Some of these mistakes affect the passive gravitational mass; others affect tidal forces in the equation of motion, etc.

The obvious mistake is that in the calculations with the equation of motion the PPN scheme begins by adopting a wrong expression for the energy density of the moving system and for the definition of the center of mass, or the "center of energy" [Nordvedt, 1968; Will, 1993, Section 6.2]. In the expression for the total energy, the gravitational energy density is written simplistically as a potential energy density $\frac{1}{2}\rho\Phi$, whereas in fact it is a combination of potential energy and field energy—the actual gravitational energy density is $\rho\Phi + \Phi_{,n}\Phi_{,n}/8\pi G$, where the first term is negative and the second term is positive. This mistake has no effect on the ratio of $M_G$ and $M_I$ for the active gravitational mass in General Relativity, where the PPN approximation gives the correct result despite this mistake (upon integration over the volume of the system, both expressions for the gravitational energy density give the same result for the net energy). However, the mistake in the treatment of gravitational energy density affects the way that the external field transfers energy and momentum to the system, and it affects the location of the center of mass and the equation of motion of a nonspherical body in an external inhomogenous gravitational field. Furthermore, the simplistic expression $\frac{1}{2}\rho\Phi$ does not include the appropriate contribution from the energy density of the Brans-Dicke scalar field. The energy density for this field consists of all the terms on the right side of Eq. (27) that involve $\phi$, except the last second-derivative term (canceled by the superpotential contribution). This energy density is proportional to the ordinary gravitational energy density [see Eq. (48)]. The failure to include the energy density in the Brans-Dicke field spoils the PPN calculations of the passive gravitational mass.

A more subtle mistake in the PPN formalism is the exclusive reliance on the conservation law $T_k^{\ n}{}_{;n} = 0$, or Eq. (1), to determine the effect of the external gravitational field on the motion. There are actually two more conservation laws within the Brans-Dicke theory. Besides the conservation law $T_k^{\ n}{}_{;n} = 0$ for nongravitational matter, there is also a conservation law for the scalar field,

$$\left[ \frac{1}{8\pi}(\phi_0 - \phi)(R_k^{\ n} - \tfrac{1}{2}\delta_k^n R) + \frac{1}{8\pi}\frac{\omega}{\phi}(\phi_{,k}\phi^{,n} - \tfrac{1}{2}\delta_k^n \phi_{,l}\phi^{,l}) + \frac{1}{8\pi}(\phi_{,k}^{\ ;n} - \delta_k^n \phi_{,l}^{\ ;l}) \right]_{;n} = 0 \quad (52)$$

This conservation law can be readily obtained by taking the covariant divergence of the field equation (27), but it can also be established independently of this field equation, in consequence of the general invariance of the part of the Lagrangian involving the scalar field [see the discussion of this point in connection with Eq. (1)]. And finally there is the conservation law for the canonical energy-momentum tensor, which includes gravitational fields, scalar field, and nongravitational matter,

$$\partial_n \mathcal{T}_k{}^n = 0 \tag{53}$$

Each of these conservation laws plays a different role. Equation (1) determines the rate at which the gravitational field transfers energy and momentum to the nongravitational matter, and Eq. (52) determines the rate of transfer to the scalar field. To find the equation of motion of a test mass, we can rely exclusively on Eq. (1). However, if the moving system includes nongravitational matter, *and* scalar fields, *and* gravitational self-fields, we need not only Eq. (1), but also Eqs. (52) and (53) to determine the complete rate of transfer of energy and momentum. In fact, we do not really need Eqs. (1) and (52) at all, because Eq. (53) includes all forms of energy and momentum and completely determines all the energy and momentum transfers across the boundaries of the system.

Thus, the conventional formulation of the PPN equation of motion (by Will and others) is clearly wrong. The PPN scheme relies exclusively on the conservation law $T_k{}^n{}_{;n} = 0$, which is superseded by the general conservation law (53). This mistake affects not only the calculation of the ratio of gravitational and inertial mass, but also other aspects of the motion, whenever the moving body is not a test mass, that is, whenever the moving body contains gravitational self-energy.

Scalar-tensor theories without general invariance (bi-metric theories, with a stipulated background geometry) are a different can of worms. The derivation in Appendix 1 does not apply to them, because this derivation hinges on general invariance and Noether's theorem. It then becomes necessary to check the concordance (or discordance) of the gravitational source tensor and the canonical energy-momentum tensor by brute force. The PPN calculations of "preferred frame" effects in such theories did not consider the distinction between the source tensor and the canonical tensor, and this means they probably underestimated the deviations from geodesic motion seen in these theories. Roughly, we can expect that the discrepancy between gravitational and inertial mass will be worse than in the Brans-Dicke theory, because, in the absence of general invariance, any agreement between the source tensor and the canonical tensor is purely coincidental. Inventors of such theories need to check their inequalities of $M_G$ and $M_I$—the burden of proof rests on their own shoulders.

**b) Vector theories.** Vector theories, with a vector field in addition to the usual gravitational tensor field, also violate the equality of gravitational and inertial mass, for much the same reason as in the scalar case—the long-range vector field modifies the asymptotic equations for $h_{00}$ and $h_{kn}$, so the gravitational and inertial masses calculated from these asymptotic expressions are different. In the case of a test mass, the difference is a harmless constant factor, which can be absorbed in the definition of the gravitational constant $G$. But in the case of a body of larger mass, the difference is of the order of magnitude of the gravitational self-energy. Will [1993, p. 185] gives a general formula for the difference between the gravitational and inertial mass in terms of the gravitational self-energy. His formula, obtained by evaluating the passive mass from the equation of motion rather than the active mass, is afflicted with the mistakes mentioned in the scalar case, and these mistakes alter the dependence on the adjustable constants of the vector theory, so the result changes quantitatively, although not qualitatively.

For a correct calculation of the gravitational vs. inertial mass in vector theories we can follow the same steps as in the Brans-Dicke case. For instance, in the so-called vector-metric theory of Will and Nordvedt or Hellings and Nordvedt [Will, 1993, Section 5.4], there is a difference between the magnitudes of the asymptotic components $h_{00}$ and $h_{11}, h_{22}, h_{33}$ of the metric tensor, and there is a difference between the source tensor and the canonical energy-momentum tensor. As in Eq. (40), this leads to a modification of the relationship between $M_G$ and $M_I$. For the calculation of the extra term in this relationship, it is necessary evaluate all the relevant energy contributions in the canonical energy-momentum tensor. The resulting difference between the gravitational and inertial masses is then approximately proportional to the gravitational self-energy, but with a constant of proportionality different from that given by Will. The calculations are tedious and not especially illuminating.

**c) Scalar-vector-tensor theory (MOG).** The scalar-vector-tensor gravity theory, also called Modified Gravity (MOG) of Moffat [Moffat, 2006; Moffat and Toth, 2008, 2009] includes several scalar and vector fields that modify the long-distance behavior of the gravitational interaction. By such modifications, the theory seeks to explain the rotation curves of galaxies and the accelerating expansion of the universe, without invoking the ad hoc hypotheses of dark matter and dark energy.

In contrast to the Brans-Dicke and other "metric" theories in which the extra scalar and vector fields couple only to the metric tensor field but not directly to ordinary (nongravitational) matter, the MOG vector field couples directly to matter. The authors of this theory claim that their vector field is "coupled universally to matter," but they specify this coupling only for the simple case of classical particles, for which they prescribe an interaction proportional to the rest mass. They give no prescription for how to couple to other matter (e.g., electromagnetic fields). In the absence of a general prescription for constructing such couplings, the theory suffers from ambiguities. For instance, the rest-mass energy of most physical particles includes electromagnetic self-energy, whenever the particle is charged (say, a proton) or contains equal positive and negative charges in its interior (say, a neutron). This means that a coupling of the MOG vector field to the rest mass implies a coupling to electromagnetic fields. But it is not at all clear that one can contrive an explicit coupling between the MOG vector field and electromagnetic fields that is effectively equivalent to a coupling to the energy of the latter fields, as required for compliance with the principle of equivalence and the Eötvös experiments.

Apart from (unknown) contributions arising from renormalizations in quantum field theory, electromagnetic field energy makes only a modest contribution to the energy content of ordinary matter, so miscouplings of electromagnetic fields might not have detectable consequences, within present experimental limits. However, the same is not true for miscouplings of gluon fields, which have field equations analogous to those of electromagnetic fields and presumably have the same coupling problems. The contribution of gluon field energy to the energy content of physical particles is large—it is the dominant contribution to the rest mass of protons and neutrons. A coupling of the MOG vector field to the rest mass of physical particles implies a coupling to gluon fields. General Relativity specifies a universal coupling of the metric tensor field to electromagnetic fields, gluon fields, and all other matter fields by the insertion of the metric tensor and covariant derivatives in the matter Lagrangian. The universality of this coupling, via the resulting energy-momentum tensor [see Eq. (6)], plays a key role in the proof of the equality of gravitational and inertial masses given in Section III. Until a corresponding universal coupling for the MOG vector is specified, the theory remains incomplete, and it is impossible to prove (or disprove) the equality of gravitational and inertial masses. On the basis of the results obtained in the Brans-Dicke theory, we expect that the gravitational and inertial masses in MOG differ by, at least, a term proportional to the gravitational self-energy, and probably also by terms involving other field self-energies, from electromagnetic fields and/or gluons fields, depending on what couplings are adopted.

A further complication of MOG lurks in its geometrical interpretation. The authors of this theory suppose that the metric of spacetime corresponds in the usual way to the tensor field, with a proper time interval $d\tau = (g_{\mu\nu}dx^{\mu}dx^{\nu})^{1/2}$. However, their action principle for the motion of particles [Moffat and Toth, 2008, Eq. (19)] suggests that the theory involves a Randers-Finsler geometry, with a proper time interval $d\tau = (g_{\mu\nu}dx^{\mu}dx^{\nu})^{1/2} + A_{\mu}u^{\mu}$, where $A_{\mu}$ is proportional to the MOG vector field. This physical interpretation of the MOG field is confirmed by an analysis of the gravitational redshift. To calculate the redshift, we can adopt Einstein's Gedankenexperiment, which simulates gravitational effects—that is, the combined effects of the (weak) gravitational field $g_{00} = 1 + h_{00}$ and the MOG vector field—by means of an elevator accelerating upward with an acceleration equal to that of free fall. The redshift of periodic signals[19] sent upward then equals the Doppler shift generated by the motion of the elevator, $\Delta\nu / \nu \approx -g\Delta z$, where $g$ is the net free-fall acceleration. In MOG, the net acceleration is $g \approx \frac{1}{2}\partial h_{00} / \partial z + \partial A_0 / \partial z$ for $u^0 \approx 1$. Thus the redshift of the signal frequency becomes $\Delta\nu / \nu \approx -g\Delta z \approx -(\frac{1}{2}\Delta h_{00} + \Delta A_0)$. This disagrees with the change of proper-time interval between successive signals if $d\tau$ is calculated according to the proper-time formula of the tensor geometry, but it agrees if $d\tau$ is calculated according to the Finsler geometry, in which $\Delta(d\tau) / d\tau \approx \Delta[(1 + h_{00})^{1/2} + A_0] \approx \frac{1}{2}\Delta h_{00} + \Delta A_0$. The Finsler geometry is an unanticipated complication in MOG, and the physical and mathematical implications need to be explored (for instance, how must the affine connection and the Riemann tensor be modified in this theory?).[20]

## APPENDIX 1: RELATION BETWEEN ENERGY-MOMENTUM TENSORS

Here, I will establish equality of the total energy and the total momentum calculated from the volume integrals of the $_\alpha{}^0$ components of the canonical energy-momentum tensor density for matter and the gravitational source tensor density:

$$\int \left( -\mathcal{L}_{\mathrm{m}} \delta_\alpha^0 + \left[ \frac{\partial \mathcal{L}_{\mathrm{m}}}{\partial Q_{i,0}} - \left( \frac{\partial \mathcal{L}_{\mathrm{m}}}{\partial Q_{i,\nu,0}} \right)_{,\nu} \right] Q_{i,\alpha} + \frac{\partial \mathcal{L}_{\mathrm{m}}}{\partial Q_{i,\nu,0}} Q_{i,\alpha,\nu} \right) d^3 x = \int \left( -2 g_{\mu\alpha} \frac{\delta \mathcal{L}_{\mathrm{m}}}{\delta g_{\mu 0}} \right) d^3 x \quad \text{(A1)}$$

The matter Lagrangian $\mathcal{L}_{\mathrm{m}}$ in this equation includes second derivatives of the fields—such as the second derivatives of metric tensor that are characteristic of nonminimal gravitational couplings—and these second derivatives contribute the $\partial \mathcal{L}_{\mathrm{m}} / \partial Q_{i,\nu,0}$ terms in the canonical tensor and corresponding terms in the source tensor, on the left and right sides of Eq. (A1), respectively.

The proof presented here is a generalization of a proof I published previously [Ohanian, 1973]. This published proof was part of a discussion of the "new, improved" energy-momentum tensor of Callan, Coleman, and Jackiw [1970], and this may have created the impression that it was limited to that particular case; furthermore, the proof rested on the assumption that the Lagrangian contained second derivatives only linearly, so these second derivatives can be eliminated by an integration by parts. Here I present a more general proof, for a Lagrangian containing second derivatives in any combination. Third and higher derivatives are excluded, but this is probably not an essential restriction (I have not explored what happens in the presence higher derivatives because the calculations are tedious and not of any practical significance; I know of attempts to use nonlinear second-derivative Lagrangians in gravitational theory—roughly, squares of the Riemann tensor—but not third derivatives).

The proof relies on an examination of the effects of performing an infinitesimal coordinate transformation in the action. This technique was used by Rosenfeld [1940] to derive identities for the source tensor and by Goldberg [1958] for the construction of the gravitational energy-momentum pseudotensor. But neither Rosenfeld nor Goldberg exploited this technique for a comparison of the corresponding canonical energy-momentum tensor density and the gravitational source tensor density of matter [the integrands on the left and right sides of Eq. (A1), respectively].[21] By direct calculation of the left side of Eq. (A1), Bergmann and Thomson [1953] confirmed that it equals the right side for the special case of the electron (spin ½), but they did not attempt a general proof for arbitrary particles with arbitrary nongravitational interactions.

The gravitational and matter Lagrangian densities in Eq. (3) are scalar densities, so that their products by the volume element $d^4 x$ are scalars. For the matter Lagrangian $\mathcal{L}_{\mathrm{m}}$ this means that under an infinitesimal general coordinate transformation $x^\mu \to x^\mu + \xi^\mu$ (regarded as an *active* transformation, that is, a bodily displacement of the physical fields by $\xi^\mu$) the transformation law is

$$\delta \mathcal{L}_{\mathrm{m}} = -(\mathcal{L}_{\mathrm{m}} \xi^{\mu})_{,\mu} \tag{A2}$$

Written explicitly in terms of the changes $\delta Q_i$ of the fields, $\delta \mathcal{L}_m$ becomes

$$\delta \mathcal{L}_{\mathrm{m}} = \left( \delta Q_i \frac{\partial}{\partial Q_i} + \delta Q_{i,\mu} \frac{\partial}{\partial Q_{i,\mu}} + \delta Q_{i,\mu,\nu} \frac{\partial}{\partial Q_{i,\mu,\nu}} \right) \mathcal{L}_{\mathrm{m}}$$

$$= \delta Q_i \frac{\delta \mathcal{L}_{\mathrm{m}}}{\delta Q_i} + \left[ \delta Q_i \frac{\partial \mathcal{L}_{\mathrm{m}}}{\partial Q_{i,\mu}} + \delta Q_{i,\nu} \frac{\partial \mathcal{L}_{\mathrm{m}}}{\partial Q_{i,\mu,\nu}} - \delta Q_i \left( \frac{\partial \mathcal{L}_{\mathrm{m}}}{\partial Q_{i,\mu,\nu}} \right)_{,\nu} \right]_{,\mu} \tag{A3}$$

where, in the second line, $\delta \mathcal{L}_{\mathrm{m}} / \delta Q_i$ is the usual variational derivative, $\delta \mathcal{L}_{\mathrm{m}} / \delta Q_i = \partial \mathcal{L}_{\mathrm{m}} / \partial Q_i - (\partial / \partial x^{\mu})(\partial \mathcal{L}_{\mathrm{m}} / \partial Q_{i,\mu}) + (\partial^2 / \partial x^{\nu} \partial x^{\mu})(\partial \mathcal{L}_{\mathrm{m}} / \partial Q_{i,\mu,\nu})$. Inserting this expression for $\delta \mathcal{L}_{\mathrm{m}}$ on the left side of Eq. (A2), we obtain Noether's identity

$$\delta J^{\mu}_{,\mu} = \delta Q_i \frac{\delta \mathcal{L}_{\mathrm{m}}}{\delta Q_i} \tag{A4}$$

with

$$\delta J^{\mu} = -\left[ \mathcal{L}_{\mathrm{m}} \xi^{\mu} + \delta Q_i \frac{\partial \mathcal{L}_{\mathrm{m}}}{\partial Q_{i,\mu}} + \delta Q_{i,\nu} \frac{\partial \mathcal{L}_{\mathrm{m}}}{\partial Q_{i,\mu,\nu}} - \delta Q_i \left( \frac{\partial \mathcal{L}_{\mathrm{m}}}{\partial Q_{i,\mu,\nu}} \right)_{,\nu} \right] \tag{A5}$$

If $Q_i$ is a matter field, the equation of motion is $\delta \mathcal{L}_{\mathrm{m}} / \delta Q_i = 0$, and hence in the sum over $i$ on the right side of Eq. (A4) we can omit contributions from the matter variables. The equation then reduces to [22]

$$\delta J^{\mu}_{,\mu} = \delta g_{\mu\nu} \frac{\delta \mathcal{L}_{\mathrm{m}}}{\delta g_{\mu\nu}} \tag{A6}$$

which we can also write

$$\delta J^{\mu}_{,\mu} = -\tfrac{1}{2} \delta g_{\mu\nu} \sqrt{-g}\, T^{\mu\nu} \tag{A7}$$

This means that the right side of our equation is related to the gravitational source tensor, and it now remains to be seen how the left side is related to the canonical tensor.

Before we reshape the left side, we want to do some reshaping of the right side. Since $g_{\mu\nu}$ is a tensor, its change $\delta g_{\mu\nu}$ under our infinitesimal coordinate transformation is

$$\delta g_{\mu\nu} = -g_{\mu\alpha} \xi^{\alpha}_{,\nu} - g_{\alpha\nu} \xi^{\alpha}_{,\mu} - g_{\mu\nu,\alpha} \xi^{\alpha} \tag{A8}$$

With this, and with the usual definition of the covariant derivative $T^{\mu\nu}_{;\nu}$, we can rewrite the right side of Eq. (A7) to obtain

$$\delta J^{\mu}{}_{,\mu} = (\sqrt{-g}\,T^{\mu\nu}g_{\mu\alpha}\xi^{\alpha})_{,\nu} - \sqrt{-g}\,T^{\mu\nu}{}_{;\nu}g_{\mu\alpha}\xi^{\alpha} \tag{A9}$$

But $T^{\mu\nu}{}_{;\nu} = 0$,[23] so this equation reduces to

$$\delta J^{\mu}{}_{,\mu} = (\sqrt{-g}\,T_{\alpha}{}^{\nu}\xi^{\alpha})_{,\nu} \tag{A10}$$

We next need an explicit expression for $\delta J^{\mu}$ in terms of $\xi^{\mu}$. The fields $Q_i$ transform according to

$$\delta Q_i = -Q_{i,\mu}\xi^{\mu} - \Lambda_{in\mu}{}^{\alpha}\xi^{\mu}{}_{,\alpha}Q_n \tag{A11}$$

and

$$\delta Q_{i,\nu} = -Q_{i,\mu\nu}\xi^{\mu} - Q_{i,\mu}\xi^{\mu}{}_{,\nu} - \Lambda_{in\mu}{}^{\alpha}\xi^{\mu}{}_{,\alpha,\nu}Q_n - \Lambda_{in\mu}{}^{\alpha}\xi^{\mu}{}_{,\alpha}Q_{n,\nu} \tag{A12}$$

where the matrices $\Lambda_{in\mu}{}^{\alpha}$ are constant and are completely determined by the Lorentz-transformation properties of the fields (for instance, for a scalar field, the $\Lambda$ matrix is zero; for a contravariant vector field, with $Q_n \equiv A^n$, $n = 0,1,2,3$, the $\Lambda$ matrix is $\Lambda_{in\mu}{}^{\alpha} = \delta_n^{\alpha}\delta_{\mu}^i$, etc.). Combining all of this, we obtain

$$-\left\{ \mathcal{L}_{\mathrm{m}}\xi^{\mu} + \left[ \frac{\partial \mathcal{L}_{\mathrm{m}}}{\partial Q_{i,\mu}} - \left( \frac{\partial \mathcal{L}_{\mathrm{m}}}{\partial Q_{i,\nu,\mu}} \right)_{,\nu} \right] (-Q_{i,\alpha}\xi^{\alpha} - \Lambda_{in\alpha}{}^{\beta}\xi^{\alpha}{}_{,\beta}Q_n) \right.$$

$$\left. + \frac{\partial \mathcal{L}_{\mathrm{m}}}{\partial Q_{i,\nu,\mu}} \left( -Q_{i,\beta,\nu}\xi^{\beta} - Q_{i,\beta}\xi^{\beta}{}_{,\nu} - \Lambda_{in\beta}{}^{\alpha}\xi^{\beta}{}_{,\alpha,\nu}Q_n - \Lambda_{in\beta}{}^{\alpha}\xi^{\beta}{}_{,\alpha}Q_{n,\nu} \right) \right\}_{,\mu} = (\sqrt{-g}\,T_{\alpha}{}^{\nu}\xi^{\alpha})_{,\nu} \tag{A13}$$

This equation must be true for every choice of the functions $\xi^{\mu}$. In particular, at any given point, we can choose $\xi^{\alpha}$ such that $\xi^{\alpha} = 0$, but the first, second, and third derivatives are not zero and take arbitrary values. The coefficients of each of these derivatives on both sides of the equation must then be equal. For the first derivative $\xi^{\alpha}{}_{,\beta}$ this implies

$$-\mathcal{L}_{\mathrm{m}}\delta_{\alpha}{}^{\beta} + \left[ \frac{\partial \mathcal{L}_{\mathrm{m}}}{\partial Q_{i,\beta}} - \left( \frac{\partial \mathcal{L}_{\mathrm{m}}}{\partial Q_{i,\nu,\beta}} \right)_{,\nu} \right] Q_{i,\alpha} + \frac{\partial \mathcal{L}_{\mathrm{m}}}{\partial Q_{i,\nu,\beta}} Q_{i,\alpha,\nu}$$

$$+\left(\frac{\partial \mathcal{L}_m}{\partial Q_{i,\nu,\beta}} Q_{i,\alpha}\right)_{,\nu} + \left\{\left[\frac{\partial \mathcal{L}_m}{\partial Q_{i,\mu}} - \left(\frac{\partial \mathcal{L}_m}{\partial Q_{i,\nu,\mu}}\right)_{\nu}\right]\Lambda_{in\alpha}{}^{\beta} Q_n\right\}_{,\mu} = \sqrt{-g}\, T_\alpha{}^{\beta} \qquad (A14)$$

The first line on the left side of this equation is the standard canonical energy-momentum tensor for a Lagrangian with second derivatives. The second line consists of a sum of divergences, and it remains to be shown that these divergences do not contribute to the conservation law for the energy-momentum tensor nor to the calculation of the total energy and momentum.

Although the second derivatives are also arbitrary, they are subject to the symmetry condition $\xi^{\beta}{}_{,\mu,\nu} = \xi^{\beta}{}_{,\nu,\mu}$. Hence the coefficient of $\xi^{\beta}{}_{,\nu,\mu}$ on the left side of Eq. (A13) must either be zero or else equal to some quantity $W_\beta{}^{[\mu\nu]}$ antisymmetric in $\mu,\nu$:

$$\left[\frac{\partial \mathcal{L}_m}{\partial Q_{i,\mu}} - \left(\frac{\partial \mathcal{L}_m}{\partial Q_{i,\alpha,\mu}}\right)_{,\alpha}\right]\Lambda_{in\beta}{}^{\nu} Q_n + \frac{\partial \mathcal{L}_m}{\partial Q_{i,\nu,\mu}} Q_{i,\beta}$$

$$+\left(\frac{\partial \mathcal{L}_m}{\partial Q_{i,\mu,\alpha}} Q_n\right)_{,\alpha}\Lambda_{in\beta}{}^{\nu} + \frac{\partial \mathcal{L}_m}{\partial Q_{i,\alpha,\mu}}\Lambda_{in\beta}{}^{\nu} Q_{n,\alpha} = W_\beta{}^{[\mu\nu]} \qquad (A15)$$

The divergence $W_\beta{}^{[\mu\nu]}{}_{,\nu}$ is a superpotential, and it identically obeys the conservation law $W_\beta{}^{[\mu\nu]}{}_{,\nu,\mu} = 0$.

Finally, we examine the terms involving the third derivatives of $\xi^\alpha$. According to Eq. (A13),

$$\frac{\partial \mathcal{L}_m}{\partial Q_{i,\nu,\mu}}\Lambda_{in\beta}{}^{\alpha} Q_n \xi^{\beta}{}_{,\alpha,\nu,\mu} = 0 \qquad (A16)$$

which implies that $(\partial \mathcal{L}_m / \partial Q_{i,\nu,\mu})\Lambda_{in\beta}{}^{\alpha} Q_n$ is antisymmetric in $\alpha,\nu$ (or, alternatively, anti-symmetric in $\alpha,\mu$; these alternatives are equivalent, because $Q_{i,\nu,\mu}$ is symmetric in $\nu,\mu$).

As a consequence of Eq. (A15), we have

$$\left(\frac{\partial \mathcal{L}_m}{\partial Q_{i,\nu,\beta}} Q_{i,\alpha}\right)_{,\nu} = -\left\{\left[\frac{\partial \mathcal{L}_m}{\partial Q_{i,\beta}} - \left(\frac{\partial \mathcal{L}_m}{\partial Q_{i,\sigma,\beta}}\right)_{,\sigma}\right]\Lambda_{in\alpha}{}^{\nu} Q_n\right\}_{,\nu}$$

$$-\left[\frac{\partial \mathcal{L}_m}{\partial Q_{i,\beta,\sigma}} Q_n \Lambda_{in\alpha}{}^{\nu}\right]_{,\sigma,\nu} - \left[\frac{\partial \mathcal{L}_m}{\partial Q_{i,\sigma,\beta}} Q_{n,\sigma}\Lambda_{in\alpha}{}^{\nu}\right]_{,\nu} + W_\alpha{}^{[\beta\nu]}{}_{,\nu} \qquad (A17)$$

Here the first term in the second line can be omitted because of the antisymmetry implied by Eq. (A16). Substituting the other terms into the second line of Eq. (A14), we obtain

$$
-\mathcal{L}_{\mathrm{m}}\delta_\alpha^{\ \beta} + \left[\frac{\partial \mathcal{L}_{\mathrm{m}}}{\partial Q_{i,\beta}} - \left(\frac{\partial \mathcal{L}_{\mathrm{m}}}{\partial Q_{i,\nu,\beta}}\right)_{,\nu}\right]Q_{i,\alpha} + \frac{\partial \mathcal{L}_{\mathrm{m}}}{\partial Q_{i,\nu,\beta}}Q_{i,\alpha,\nu}
$$

$$
-\left\{\left[\frac{\partial \mathcal{L}_{\mathrm{m}}}{\partial Q_{i,\beta}} - \left(\frac{\partial \mathcal{L}_{\mathrm{m}}}{\partial Q_{i,\sigma,\beta}}\right)_{,\sigma}\right]\Lambda_{in\alpha}^{\ \nu}Q_n \right\}_{,\nu} - \left[\frac{\partial \mathcal{L}_{\mathrm{m}}}{\partial Q_{i,\sigma,\beta}}Q_{n,\sigma}\Lambda_{in\alpha}^{\ \nu}\right]_{,\nu} + W_\alpha^{\ [\beta\nu]}_{\ ,\nu}
$$

$$
+\left\{\left[\frac{\partial \mathcal{L}_{\mathrm{m}}}{\partial Q_{i,\nu}} - \left(\frac{\partial \mathcal{L}_{\mathrm{m}}}{\partial Q_{i,\mu,\nu}}\right)_{\mu}\right]\Lambda_{in\alpha}^{\ \beta}Q_n\right\}_{,\nu} = \sqrt{-g}\,T_\alpha^{\ \beta} \qquad (A18)
$$

The two terms with braces { } in this equation are divergences that combine to form a single divergence of the same kind as $W_\alpha^{\ [\beta\nu]}_{\ ,\nu}$, with the same kind of $\beta,\nu$ asymmetry; therefore these terms can be subsumed in $W_\alpha^{\ [\beta\nu]}_{\ ,\nu}$. The remaining term $[(\partial \mathcal{L}_{\mathrm{m}}/\partial Q_{i,\sigma,\beta})Q_{n,\sigma}\Lambda_{in\alpha}^{\ \nu}]_{,\nu}$ also has the same kind of $\beta,\nu$ asymmetry, as can be seen by comparing this term with the term $(\partial \mathcal{L}_{\mathrm{m}}/\partial Q_{i,\sigma,\beta})Q_n\Lambda_{in\alpha}^{\ \nu}$ discussed in connection with Eq. (A16)—the $\beta,\nu$ asymmetry of this term is not affected by the differentiation of $Q_n$ with respect to $x^\sigma$. Therefore the remaining term is again a divergence with the same kind of $\beta,\nu$ asymmetry and it can again be subsumed in $W_\alpha^{\ [\beta\nu]}_{\ ,\nu}$. Our final result for the relationship between the canonical energy-momentum tensor and the gravitational source tensor is then

$$
-\mathcal{L}_{\mathrm{m}}\delta_\alpha^{\ \beta} + \left[\frac{\partial \mathcal{L}_{\mathrm{m}}}{\partial Q_{i,\beta}} - \left(\frac{\partial \mathcal{L}_{\mathrm{m}}}{\partial Q_{i,\nu,\beta}}\right)_{,\nu}\right]Q_{i,\alpha} + \frac{\partial \mathcal{L}_{\mathrm{m}}}{\partial Q_{i,\nu,\beta}}Q_{i,\alpha,\nu} + W_\alpha^{\ [\beta\nu]}_{\ ,\nu} = \sqrt{-g}\,T_\alpha^{\ \beta} \qquad (A19)
$$

The term $W_\alpha^{\ [\beta\nu]}_{\ ,\nu}$ indicates a discrepancy between these two tensors. However, because of the $\beta,\nu$ asymmetry, this term identically obeys the conservation law $W_\alpha^{\ [\beta\nu]}_{\ ,\nu,\beta} \equiv 0$. Furthermore, under the assumption the matter is confined to a finite volume (or its energy and momentum densities tend to zero sufficiently strongly for $r \to \infty$, as in the case of an electrostatic field), then the space integrals of the $_\alpha^{\ 0}$ components of the canonical and source tensors will be equal. With $\beta = 0$, $W_\alpha^{\ [\beta\nu]}_{\ ,\nu}$ reduces to $W_\alpha^{\ [0k]}_{\ ,k}$, which is a spatial divergence and can be converted into a surface integral over a surface at large distance where the matter fields are zero (or tend to zero), so $W_\alpha^{\ [0k]} \to 0$ and therefore the surface integral is zero. The term $W_\alpha^{\ [\beta\nu]}_{\ ,\nu}$ in Eq. (A19) can then be ignored in the calculation of the total energy and momentum, and, accordingly, we obtain the result stated in Eq. (A1).

A few final comments: Eq. (A11) asserts that all of the fields $Q_i$ belongs to some representation of the Lorentz group (with Lorentz-transformation coefficients $\Lambda_{in\mu}{}^{\alpha}$). That this must be true for quantum-mechanical fields was established long ago by Wigner in one of his seminal papers [Wigner, 1939]. A similar argument can be made for classical fields by noting that if a field with components $Q_i$ describes the state of a system in one inertial reference frame, then in another reference frame the state of the system must necessarily be described by a linear superposition of the components $Q_i$, at least when the field is weak (small $Q_i$) or when the change of reference frame is small [as in Eq. (A11)], so a Taylor-series expansion is applicable. This linear superposition ensures that such weak fields or small changes in fields constitute representations of the Lorentz group.

The construction of the canonical energy-momentum tensor by examination of infinitesimal translations can also be applied to the case of the gravitational field. According to Goldberg, this yields the usual canonical energy-momentum tensor adopted by Einstein (who imitated Born's construction of the canonical energy momentum, published several years earlier in work that was probably known to Einstein). However, in his construction of this tensor, Goldberg commits a slight misstep: instead of using the scalar density $R\sqrt{-g}$ as starting point for the canonical energy-momentum tensor, he uses the truncated expression $\mathcal{L}_G$, which is *not* a scalar density. If we use the proper scalar density $R\sqrt{-g}$, then the canonical energy-momentum tensor includes an extra term $(-S^{\nu}\delta_{\alpha}^{\beta}+S^{\beta}\delta_{\alpha}^{\nu})_{,\nu}$ with second derivatives of the gravitational field [Ohanian, 1973]. This extra term arises from the divergence $S^{\mu}{}_{,\mu}$ in Eq. (3), and it leads to a discrepancy of the same kind as $W_{\alpha}{}^{[\beta\nu]}{}_{,\nu}$ in Eq. (A19). However, again, this discrepancy does not affect the conservation law nor the calculated values of the total energy and momentum.

For purposes quantum theory or cosmology we might want to add terms to the Lagrangian that are nonlinear functions of the Riemann tensor (for instance, $R^2\sqrt{-g}$, or $R^{\mu\nu}R_{\mu\nu}\sqrt{-g}$, etc.). As already mentioned in connection with Eq. (2), such terms can be included in $\mathcal{L}_m$. Formally, such terms then produce extra gravitational sources in addition to the source $T_{\alpha}{}^{\beta}$ produced by matter. The strength of these sources typically decreases as $1/r^6$; there will then be no trouble from the volume integral of the corresponding $W_{\alpha}{}^{[0\nu]}{}_{,\nu}$ term, and an equation analogous to Eq. (A1) will again be valid. So the equality of inertial and gravitational masses will remain valid.

## ENDNOTES

[1] Imposition of a size limit does not necessarily preserve validity of SEP. For instance, a nonspherical gyroscope in free fall in a gravitational field will precess in response to torques exerted by tidal forces (the equinoctial precession of the axis of the Earth is an example of such a precession). The precession rate depends on the shape of the gyro-scope, but not on its size. This violation of SEP can be eliminated by placing a short time limit on the measurement, so that the angle of precession becomes unobservable.

[2] Sometimes the conservation law (1) is said to be a consequence of Einstein's field equations $R_\mu{}^\nu - \frac{1}{2}\delta_\mu^\nu R = 16\pi G T_\mu{}^\nu$ and the Bianchi identity $(R_\mu{}^\nu - \frac{1}{2}\delta_\mu^\nu R)_{;\nu} = 0$. But Eq. (1) is actually independent of that, because of the general-invariance symmetry of the Lagrangian, which directly implies $T^{\mu\nu}{}_{;\nu} = 0$, regardless of the Einstein field equations. See endnote 23 or see Landau and Lifshitz [1962, Section 94].

[3] It is possible to convert Eq. (1) into the geodesic equation by inserting delta functions for the energy momentum tensor of the particle (see e.g., Hobson et al., 2006, Section 8.8). This is a pretty piece of mathematics, but bad physics—a delta-function mass distribution generates a divergent gravitational self-field and divergent gravitational self-energy, which will severely modify the external gravitational field and derail any attempt to deduce the geodesic equation. In the 1940s Einstein and Bergmann and, later, Infeld and Hoffmann, claimed they could deduce the motion of pointlike singularities in the gravitational field and deduce the geodesic equation by considering the modifications of the external gravitational field. But the real meaning of these results is obscure, because Einstein had no understanding of the geometry of singularities and their horizons in General Relativity; and furthermore, any singularity smaller than the Planck mass ($10^{-5}$ g) or the Planck size ($10^{-33}$ cm) is subject to drastic quantum effects and cannot be treated by the purely classical calculations of Einstein and his associates.

[4] Landau and Lifshitz [1962, p. 312] went so far as to argue that examination of the case of electromagnetic fields suffices to establish the concordance of the source and canonical tensors in general. They contended that because both of these tensors are conserved in the local inertial reference frame, they always must coincide except for a factor of proportionality, which can be determined by merely checking one particular case (such as the electromagnetic case). This argument is nonsense. The tensors can differ by a term that has zero divergence, such as the term $W_\alpha^{[\beta\nu]}{}_{,\nu}$.

[5] Whether a special proof is required for spinor fields is unclear to me. It would seem that all that is required to generalize my proof to spinor fields is to adopt vierbein fields as the gravitational field variables, so the variational derivative $\delta/\delta g_{\mu\nu}$ needed to calculate the variation of the Lagrangian with respect to $g_{\mu\nu}$ is replaced by the variational derivative $\delta^2/\delta V_\mu \delta V_\nu$. The vierbein variables $V_\mu$ can then be included among the field variables $Q_i$ in the equations in Appendix 1.

[6] This field equation implies the conservation law $\partial_\nu(\sqrt{-g}\, t_\mu{}^\nu + \sqrt{-g}\, T_\mu{}^\nu) = 0$, which can also be derived from the more familiar conservation law (1), by using the field equation to eliminate $T_\alpha{}^\mu$ in the first term on the right side of Eq. (1).

[7] We use a script letter $\mathcal{T}_\mu{}^\nu$ for the combined energy-momentum tensor of matter and gravitation. In contrast, Tolman [1934] uses this script letter for the energy-momentum tensor of matter.

[8] A distinctive characteristic of the energy defined by the canonical construction is that the energy operator (that is, the Hamiltonian $H$) is the generator of time translations for the field operators and their canonically conjugate momenta [Goldstein, 1959, p. 364; Bergmann, 1985; Schweber, 1961, p. 193; Weinberg, 1995, vol. 1, p. 311]. In quantum mechanics, in the Heisenberg picture, this means that the time derivative of any function $F$ of the field operators and their canonical momenta is given by the commutator of $H$ and $F$, $\partial F/\partial t = [F, H]/i\hbar$, which shows that $H$ generates an increment in $F$ per unit time [in the Schrödinger picture, this equation corresponds to the equation for the time derivative of the state vector, $\partial\Psi/\partial t = (1/i\hbar)H\Psi$]. The identification of the Hamiltonian with the generator of time translations is also valid in classical mechanics, because of an analogous formula involving the Poisson bracket instead of the commutator, $\partial F/\partial t = \{F, H\}$.

Another important property of the canonical energy-momentum tensor is that under a change of field variables, say from $Q_i$ to $U_j = U_j(Q_i)$, this tensor remains invariant. Under such a redefinition of the field variables, the Lagrangian density $\mathcal{L}$ is invariant in content, although not in form, $\mathcal{L}(U_j) = \mathcal{L}(Q_i(U_j))$, and the energy momentum tensor is then invariant by the chain rule for derivatives, which yields $(Q_{i,\mu}\partial\mathcal{L}/\partial Q_{i,\nu} - \delta_\mu^\nu\mathcal{L}) = (U_{i,\mu}\partial\mathcal{L}/\partial U_{i,\nu} - \delta_\mu^\nu\mathcal{L})$. Obviously, this invariance is required for the consistent interpretation of the canonical tensor as physical energy and momentum, because the latter cannot depend on how we elect to describe the fields.

[9] This formula was first given by Komar [1959], who presented it as an expression for the Schwarzschild mass. He did not distinguish between gravitational and inertial masses and implicitly took it for granted that they would be equal.

[10] Another neat formula for the mass, valid for static configurations of spherical symmetry, is $M_I = \int \rho*(r)4\pi r^2 dr$, where $\rho*$ is the density of mass in the local inertial reference frame, and $r$ is the radial Schwarzschild coordinate. Note that $4\pi r^2 dr$ is not the actual measured volume element, but merely a coordinate volume element, which is actually smaller than the actual measured volume element $4\pi r^2 dr\sqrt{-g_{11}}$. Thus, $M_I$ is smaller than the sum of local mass densities, as expected [Harrison et al., 1965, p. 14].

[11] Equation (7) is applicable to the empty space Schwarzschild solution, provided we adopt coordinates that are free of singularities on the hypersurface of integration. However, the Tolman formula (23) is not applicable, because it hinges on a time-independent gravitational field, whereas the interior Schwarzschild solution is time dependent.

セ

[12] Wald [1984, p. 318] argues, with good reason, that the time-like singularity of the ideal Kerr-Newman solution is actually preempted by a light-like Cauchy horizon that cuts the solution off just as it reaches the beginning of the time-like singularity. But in any case, we have available a spacelike hypersurface free of singularities.

[13] Momentum conservation could conceivably be violated in some alternative theories of gravitation, but it cannot be violated in General Relativity, which is a theory derivable from a Lagrangian, and which therefore obeys the energy-momentum conservation laws.

[14] For a system with strong gravitational fields, but without a horizon, Niewenhuizen's expression for the inertial mass is $\int T_0^{\ 0} \sqrt{-g}\, d^3 x$, which fails to include the energy density in the gravitational field [see Eq. (12)].

[15] This can be shown to be a consequence of the conservation law $T_k^{\ n}{}_{,n} = 0$, $n = 1, 2, 3$, which holds for a static system or for the time average of a periodic system. Multiplication of this conservation law by $x^l$, integration over the entire volume of the system, and integration by parts immediately yields $0 = \int x^l T_k^{\ n}{}_{,n} d^3 x = -\int T_l^{\ n} d^3 x$.

[16] In Eq. (32) of the Brans-Dicke paper, the $-C$ in the expression for $\phi$ should be $+C$.

[17] However, in proceeding to an approximate evaluation of Eq. (40) to order $M^2$, I rashly used an expression for $\int T \sqrt{-g}\, d^3 x$ derived by Tolman [1934], which is valid in General Relativity, but not in the Brans-Dicke theory. This led to a wrong final result for the approximation for the ratio of gravitational and inertial mass. Here, I will correct this mistake.

[18] The canonical tensor, or Einstein pseudo-tensor, differs from the Landau-Lifshitz tensor. Goldberg has argued that the latter—though it has the desirable feature of being symmetrical—is defective in that it its integration does not yield a four-vector energy-momentum (like that of a particle), but a four-vector density [Goldberg, 1958]. He has proposed alternative symmetric tensors free of this defect.

[19] The signals can be periodic light waves, or periodically emitted canon balls, or signal rockets, etc. The only crucial requirement is that the successive signals must have congruent wordlines.

[20] For attempts to solve the dark mass and dark matter problems by means of Finsler geometries, see Chang and Li [2008, 2009].

[21] For Lagrangians with only first derivatives, Rosenfeld comes close to obtaining my identity (A19), but he insists on formulating his equations in terms of covariant derivatives. This is a tactical blunder, because it makes the physical interpretation of conservation laws murky [see, e.g., Eq. (1)], and it ruins the crucial identity $W_\alpha^{\ [\beta\nu]}{}_{,\nu,\beta} \equiv 0$, which fails when the ordinary derivatives are replaced by covariant derivatives, which do not commute in a curved spacetime. Rosenfeld recognizes this problem toward the end of his paper, and he sidesteps it by restricting his remaining calculations to flat spacetime. [I owe thanks to Friedrich Hehl for bringing Rosenfeld's work to my attention.]

[22] If we were dealing with the complete Lagrangian, the right side of this equation would be the field equation for the gravitational field, which would be zero. We would then obtain a conservation law, according to the usual rule that symmetries imply conservation

laws (actually, we would obtain a large variety of conservation laws, because *each* conceivable choice of the function $\xi^\mu$ leads to a different conservation law [Bergmann, 1958]). The conservation law for the canonical energy-momentum tensor would emerge by selecting $\xi^\mu = \text{constant}$, but otherwise arbitrary. We can see this by writing an equation analogous to (A13) for the complete Lagrangian, with zero on the right side, and then substituting (A18). By using only one part of the complete Lagrangian in Eq. (A4), we obtain identities for the energy-momentum tensor, instead of conservation laws. [23] This covariant conservation law can be deduced directly from (A9) by integrating both sides of this equation over some arbitrary region of spacetime, and adopting a value of $\xi^\alpha = 0$ on the boundary (but nonzero in the interior). The 4-D divergence terms can then be transformed into surface terms, which vanish, and the only remaining term is $\int T^{\mu\nu}{}_{;\nu} g_{\mu\alpha}\xi^\alpha \sqrt{-g}\,d^4x = 0$. Since $\xi^\alpha$ is arbitrary within the region, $T_\alpha{}^\nu{}_{;\nu}$ must be zero [Landau and Lifshitz, 1964, Section 94]. A similar method can be used to deduce the Bianchi identity, by taking as starting point an equation analogous to (A2) for the gravitational part of the Lagrangian, instead of the matter part.

## REFERENCES

Arnowitt, R., Deser, S., and Misner, C. W., "The Dynamics of General Relativity," in Witten, L., ed., *Gravitation, an Introduction to Current Research* (Wiley & Sons, New York, 1962).

Bartlett, D. F., and Van Buren, D., Phys. Rev. Lett. **57**, 21 (1986).

Bergmann, P. G., Phys. Rev. **112**, 287 (1958).

Bergmann, P. G., and Thomson, R., Phys. Rev. **89**, 400 (1953).

Brans, C., and Dicke, R. H., Phys. Rev. **124**, 925 (1961).

Callan, C. G., Coleman, S., and Jackiw, R., Ann. Phys. **59**, 42 (1970).

Chang, Z., and Li, X., arXiv:0806.2184 and Phys. Lett. B **668**, 453 (2008).

Chang, Z., and Li, X., arXiv:0901.1023 and Phys. Lett. B **676**, 173 (2009).

Dicke, R., Phys. Rev. **126**, 1580 (1962).

Eddington, A. S., *The Mathematical Theory of Relativity* (Cambridge University Press, Cambridge, 1924).

Gamboa-Saravi, R. E., arXiv:math-ph/0306020v1 and J. Phys. **A37**, 9573 (2004).

Goldberg, J. N., Phys. Rev. **111**, 315 (1958).

Goldstein, H., *Classical Mechanics* (Addison-Wesley, Reading, MA, 1959).

Good, M. L., Phys. Rev. **121**, 311 (1961).

Harrison, B. K., Thorne, K. S., Wakano, M., and Wheeler, J. A. W., *Gravitation Theory and Gravitational Collapse* (University of Chicago Press, Chicago, 1965).

Hobson, M. P., Efstathiou, G. P., and Lasenby, A. N., *General Relativity, An Introduction for Physicists* (Cambridge University Press, Cambridge, 2006).

Komar, A., Phys. Rev. **113**, 934 (1959).

Kreuzer, L. B., Phys. Rev. **169**, 1007 (1968).

Landau, L. D., and Lifshitz, E. M., *The Classical Theory of Fields* (Addison-Wesley, Reading MA, 1962).

Moffat, J. W., arXiv:gr-qc/0506021 and J. Cosmology and Astroparticle Phys. **2006**, 004 (2006).

Moffat, J. W., and Toth, V. T., arXiv:0805.4774v3, (2008).

Moffat, J. W., and Toth, V. T., arXiv:0712.1796 [gr-qc] and Class. Quant. Grav. **26**, 085002 (2009).

Møller, C., *The Theory of Relativity* (Clarendon Press Oxford, 1952).

Nieuwenhuizen, Th. M., "Nonequivalence of Inertial Mass and Active Gravitational Mass," in Adenier, G., Fuchs, C. A., Khrennikov, A. Yu., eds., *Foundations of Probability and Physics-4* (American Institute of Physics, 2007).

Nordvedt, K., Phys. Rev. **169**, 1017 (1968).

Ohanian, H. C., Ann. Phys. **67**, 648 (1971).

Ohanian, H. C., Int. J. Theoret. Phys. **4**, 273 (1971).

Ohanian, H. C., J. Math. Phys. **14**, 1892 (1973).

Ohanian, H. C., Am. J. Phys. **45**, 903 (1977).

Padmanabhan, T., *Gravitation* (Cambridge University Press, Cambridge, 2010).

Rosenfeld, L., Mem. Acad. Roy. de Belgique **18**, 1 (1940). Reprinted in R. S. Cohen and J. J. Stachel, *Selected Papers of Léon Rosenfeld* (Reidel, Dordrecht, 1979).

Schweber, S. S., *An Introduction to Relativistic Quantum Field Theory* (Harper & Row, New York, 1961).

Soper, D. E., *Classical Field Theory* (Wiley & Sons, New York, 1972).

Tolman, R. C., *Relativity, Thermodynamics, and Cosmology* (Clarendon Press, Oxford, 1934).

Wald, R. M., *General Relativity* (University of Chicago Press, Chicago, 1984).

Weinberg, S., *Gravitation and Cosmology: Principles and Applications of the General Theory of Relativity* (Wiley & Sons, New York, 1972).

Weinberg, S., *The Quantum Theory of Fields* (Cambridge University Press, Cambridge, 1995).

Wigner, E., Ann. Math. **40**, 149 (1939); reprinted in F. J. Dyson, ed., *Symmetry Groups in Nuclear and Particle Physics* (Benjamin, New York, 1966).

Will, C. M., *Theory and Experiment in Gravitational Physics* (Cambridge University Press, Cambridge, 1993).

Will, C. M. (2005), "The Confrontation between General Relativity and Experiment," http://www.livingreviews.org/lrr-2006-3 (accessed 26 July, 2010).

Zatzkis, H., Phys. Rev. **81**, 1023 (1951).

Zhang, H.-B., arXiv:math-ph/0412064 and Comm. in Theoretical Physics **44**, 1007 (2005).